\documentclass{aastex61}

\received{November 5, 2020}
\revised{December 25, 2020}
\accepted{January 7, 2021}
\submitjournal{ApJL}

\shorttitle{The Astrophysical Journal Letters}
\shortauthors{Yan et al.}

\usepackage{amssymb,amsmath}
\usepackage{float}
\usepackage{graphicx}
\begin{document}
\title{Atmosphere escape inferred from modelling the H$\alpha$ transmission spectrum of WASP-121b}

\correspondingauthor{Jianheng Guo, guojh@ynao.ac.cn}

\author{Dongdong Yan}
\affil{Yunnan Observatories, Chinese Academy of Sciences, P.O. Box 110, Kunming 650011, People's Republic of China}
\affiliation{School of Astronomy and Space Science, University of Chinese Academy of Sciences, Beijing, People's Republic of China}
\affiliation{Key Laboratory for the Structure and Evolution of Celestial Objects, CAS, Kunming 650011, People's Republic of China}

\author{Jianheng Guo}
\affil{Yunnan Observatories, Chinese Academy of Sciences, P.O. Box 110, Kunming 650011, People's Republic of China}
\affiliation{School of Astronomy and Space Science, University of Chinese Academy of Sciences, Beijing, People's Republic of China}
\affiliation{Key Laboratory for the Structure and Evolution of Celestial Objects, CAS, Kunming 650011, People's Republic of China}

\author{Chenliang Huang}
\affil{Lunar and Planetary Laboratory, University of Arizona, Tucson, AZ 85721, US}

\author{Lei Xing}
\affil{Yunnan Observatories, Chinese Academy of Sciences, P.O. Box 110, Kunming 650011, People's Republic of China}
\affiliation{School of Astronomy and Space Science, University of Chinese Academy of Sciences, Beijing, People's Republic of China}
\affiliation{Key Laboratory for the Structure and Evolution of Celestial Objects, CAS, Kunming 650011, People's Republic of China}



\begin{abstract}
The escaping atmospheres of hydrogen driven by stellar X-ray and extreme Ultraviolet (XUV) have been detected around some exoplanets by the excess absorption of Ly$\alpha$ in far ultraviolet band. In the optical band the excess absorption of H$\alpha$ is also found by the ground-based instruments. However, it is not certain so far if the escape of the atmosphere driven by XUV can result in such absorption. Here we present the XUV driven hydrodynamic simulation coupled with the calculation of detailed level population and the process of radiative transfer for WASP-121b. Our fiducial model predicts a mass loss rate of $\sim$1.28$\times$10$^{12}$g/s for WASP-121b. Due to the high temperature and Ly$\alpha$ intensity predicted by the fiducial model, many hydrogen atoms are populated into the first excited state. As a consequence, the transmission spectrum of H$\alpha$ simulated by our model is broadly consistent with the observation. Comparing with
the absorption of H$\alpha$ at different observation times, the stellar XUV emission varies in the range of 0.5-1.5 times fiducial value, which may reflect the variation of the stellar activity. Finally, we find that the supersonic regions of the planetary wind contribute a prominent portion to the absorption of H$\alpha$ by comparing the equivalent width of H$\alpha$, which hints that a transonic outflow of the upper atmosphere driven by XUV irradiation of the host star can be detected by the ground-based telescope and the H$\alpha$ can be a good indicator of escaping atmosphere.
\end{abstract}


\section{Introduction} \label{sec:intro}

Planetary atmosphere is the key interacting layer between planets and their host stars. For close-in planets, their atmospheres can be photoevaporated by the strong irradiation of the host stars \citep{2018MNRAS.481.5315S,2019ApJ...880...90Y,2020MNRAS.491.3435S}. The escape of the atmosphere can potentially modify the atmospheric composition and structure and further influence the planetary evolution and distribution \citep{2013ApJ...775..105O,2015ApJ...808..150H,2017AJ....154..109F}. Transmission spectroscopy is a powerful method
for the study of the atmosphere, the spectra detected by which provide a lot of physical and chemical information of
the planetary atmosphere. Many atoms and molecules, such as H, He, Na, Mg, Fe, and H$_2$O, have been detected in the planetary atmosphere by this method \citep{2003Nature...422..143,2010A&A...514..72,2012ApJ...751..86,2015Nature...522..459,2007Natur.448..169T,2018Natur.557...68S,2018Sci...362.1384A,2019A&A...621A..74A,2020A&A...641L...7S,2020A&A...641A.123H,2020A&A...638A..26S}, which is crucial for studying the origin of planets.

By detecting the excess absorption of Ly$\alpha$,  the extended hydrogen atmosphere has been found in HD 209458b, HD189733b and GJ 436b\citep{2003Nature...422..143,2010A&A...514..72,2012ApJ...751..86,2015Nature...522..459}.
The Ly$\alpha$ absorption is caused by the hydrogen atoms in the ground state. Thus the absorption depth is relatively higher compared to the line absorption caused by the hydrogen atoms in the excited states. However, the Ly$\alpha$ line can be quenched by the hydrogen in the ISM and also affected by the geoemissions, which limits its observations to the space. Fortunately, the transmission spectroscopy in the optical band provides an alternative way to study the atmosphere from the ground. \cite{2007Natur.445..511B} firstly reported the detection of Balmer edge absorption in HD 209458b, even though the absorption was relatively low. After that, the H$\alpha$ transmission spectroscopy springs up for detecting hydrogen atmosphere around the exoplanets. 
So far, the excess absorption of H$\alpha$ have been observed in seven exoplanet systems (HD 189733b, KELT-9b, KELT-20b, WASP-12b, WASP-52b, WASP-121b and WASP-33b \citep{2012ApJ...751..86,2018NatAs...2..714Y,2019AJ....157...69C,2018A&A...616A.151C, 2018AJ....156..154J,2020A&A...635A.171C,2020MNRAS.494..363C,2020arXiv201101245B,2020A&A...638A..87W,2020arXiv201002118C,2020arXiv201107888Y}). All these detections of H$\alpha$ show the existence of hydrogen in the excited state, although there are some controversies over the interpretations of the H$\alpha$ signal in HD 189733b \citep{2016MNRAS.462.1012B,2017AJ....153..185C,2017AJ....153..217C}.

Among the seven systems, different assumptions are applied in explaining the H$\alpha$ signals. For HD 189733b, which has a high gravitational potential, \citet{2013ApJ...772..144C, 2017ApJ...851..150H} applied a hydrostatic model to fit the excess absorption. However, the hydrostatic assumption can not be applicable for planets with an expanding atmosphere and a low mean density. For KELT-9b, it orbits a hot A-type star at 0.03368  AU \citep{2019A&A...631A..34B}. Its equilibrium temperature is higher than 4000 K so that many hydrogen atoms are in the first excited state (H(2)). In this situation, the intense near-ultraviolet (NUV) irradiation from its host star is the main source of heating in the atmosphere \citep{2019ApJ...884L..43G}. Although for hot stars, the irradiation of the NUV could be dominant in driving the escape of the atmosphere, the emission of the XUV should be more intense than that of NUV for late-type stars \citep{2019ApJ...884L..43G}.

Therefore, this motivates us to explore the possibility that the H$\alpha$ transmission spectra are the signals of the escaping atmosphere driven by XUV. To this end, WASP-121b is an excellent target to study. Its mass and radius are 1.183 M$_J$ and 1.865 R$_J$, respectively \citep{2016MNRAS.458.4025D}. Therefore, one can expect a relatively expanding atmosphere owing to its low mean density and gravitational potential. In addition, WASP-121b is a hot Jupiter orbiting around a F6V star at a distance of 0.02544 AU. This means that the XUV irradiation received by the planet is about 1500 times higher than that received at 1 AU.
Using ESO-HARPS, \citet{2020MNRAS.494..363C} found that in the mid-transit the absorption of H$\alpha$ at the line center was about 1.87\%, along with a 5.82 km/s red-shift.  Subsequently, \citet{2020arXiv201101245B} detected about 1.4\% and 1.7\% absorption depth of H$\alpha$ for 1-UT and 4-UT with ESO-ESPRESSO, respectively, both of which along with a blue-shift.
In addition, other species such as Fe I, Fe II, MgII, and H$_2$O have also been found in its atmosphere \citep{2020MNRAS.493.2215G,2020A&A...641A.123H,2019AJ....158...91S,2020MNRAS.496.1638M, 2016ApJ...822L...4E}. An expanding and potentially escaping hydrogen has been invoked to explain the H$\alpha$ absorption \citep{2020MNRAS.494..363C,2020arXiv201101245B}. However, a detailed model is still absent in explaining the excess absorption of H$\alpha$. Thus, it is not clear if the absorption in H$\alpha$ can be attributed to an escaping atmosphere driven by XUV irradiation of the host star. In this paper, our aim is to model the H$\alpha$ transmission spectrum of WASP-121b. In Section 2, we describe the method. In Section 3, we display the results. In Section 4, we discuss the comparison with observations. In Section 5, we summarize the work and state our conclusions.

\section{Method} \label{sec:memo}

\subsection{Hydrodynamic atmosphere model} \label{subsec:model-1}

We used the hydrodynamic model \citep{2019ApJ...880...90Y} to simulate the atmospheric structure of WASP-121b and obtained the atmospheric temperature, velocity and particle number densities. The planetary and stellar parameters are based on the observations \citep{2016MNRAS.458.4025D,2020MNRAS.494..363C}. The equilibrium temperature (T$_{eq}$) is 2361 K, which is also the temperature at the bottom boundary in our model. The high temperature hints at the dissociation of H$_{2}$. The chemical composition of WASP-121b is assumed to be the same as that of WASP-121, which is calculated by the solar abundance that is modified by [Fe/H] = 0.13 \citep{2016MNRAS.458.4025D}. The integrated flux in XUV band is an important input in the simulations. Due to the lack of the stellar XUV observations, we used the age-luminosity relation \citep{2011A&A...532A...6S} to calculate the $F_{XUV}$ received by the planet. WASP-121b is about 1.5 Gigayear (Gyr), so the $F_{XUV}$ is about 37387 erg/cm$^2$/s at the orbital distance. In our calculation the value is divided by a factor of 4, which accounts for the uniform redistribution of the stellar radiation energy around the planet. Finally, the XUV SED is obtained by the XSPEC-APEC software \citep{1996ASPC..101...17A}. In the simulations the pressure at the bottom boundary is 1 $\mu$bar. The upper boundary is 7.6 R$_p$, which covers the radius of the host star. 
The above input values are called the fiducial ones in this paper. In general, the escaping models assume that the photons of Ly$\alpha$ can freely escape from the atmosphere \citep{2009ApJ...693...23M}. In the process of resonant scattering, the number of scattering that a Ly$\alpha$ photon takes to escape the atmosphere is comparable to its line center optical depth $\tau$. In the atmosphere above $\sim$ 1.1-1.2 R$_p$ where Ly$\alpha$ cooling is most efficient, $\tau \ll 1/p_{abs} $, where $p_{abs}$ is the Ly$\alpha$ photon destroy probability per scattering \citep{2017ApJ...851..150H}. Thus, the Ly$\alpha$ cooling is included in the simulations. Furthermore, the stellar tidal force is also considered in the model.

\subsection{Hydrogen populations in the excited state.}

The H$\alpha$ absorption is caused by the hydrogen atoms in the first excited state (n=2, where n is the principal quantum number). Because of the coupling of spin and orbital angular momentum, this state is split to 2s and 2p substate. In the upper thermosphere, both the collisions among particles and the radiation process affect the population of H(2). Thus, we used a non-local thermal equilibrium (NLTE) scheme to calculate the hydrogen populations based on the hydrodynamic results.
Assuming that the atmosphere is in a stationary state, the production rates of H(2) are equal to their loss rates. We find that the Ly$\alpha$ mean intensity ($\bar{J}_{Ly\alpha}$) is dominant in determining the number density of H(2p) (as shown by Equation (1)), which is consistent with that of \citet{2013ApJ...772..144C} and \citet{2017ApJ...851..150H}. Therefore, it can be approximately expressed as:
\begin{equation}
 n_{2p} \approx \frac{ B_{1s\rightarrow2p}\bar{J}_{Ly\alpha}}{A_{2p\rightarrow1s}}n_{1s}
\end{equation}
where n(1s) and n(2p) are the number densities of H(1s) and H(2p), $B_{1s\rightarrow2p}$ and ${A_{2p\rightarrow1s}}$ are the Einstein coefficients \citep{Rybicki 2004}.
The number density of H(2s) is solved in the meantime. The source of H(2s) is mainly from $\textit{l}$-mixing; and the sink of H(2s) is mostly due to the $\textit{l}$-mixing ($\textless$ 2.5 R$_p$) and photoionization ($\textgreater$ 2.5 R$_p$). Therefore, it can be approximately expressed as:
\begin{equation}
 n_{2s} \approx \frac{ C_{2p\rightarrow2s(p) }n_{2p}n_{p} + C_{2p\rightarrow2s(e) }n_{2p}n_{e} }{C_{2s\rightarrow2p(p) }n_{p} + C_{2s\rightarrow2p(e) }n_{e} +\Gamma_{2s}}
\end{equation}
where $C_{2p\leftrightarrow2s(p)}$ and $C_{2p\leftrightarrow2s(e)}$ are the protons' and electrons' collisional transition rates \citep{1955PPSA...68..457S} between 2p and 2s, $n_{p}$ and  $n_{e}$ are the number densities of protons and electrons, and $\Gamma_{2s}$ is the photoionization rate of H(2s). Although n$_{2p}$ and n$_{2s}$ can be estimated by Equation (1) and (2), in the simulations we solved the equation of rate equilibrium for H(2p) and H(2s) by including the reactions of radiative excitation and de-excitation, collision excitation and de-excitation, photoionization and recombination, etc. The equations of rate equilibrium are the same to that of \citet{2017ApJ...851..150H}.

Because there is currently no available Ly$\alpha$ profile of the host star, in our model we took the Ly$\alpha$ flux of $\zeta$ Dor \citep{2013ApJ...766...69L} to replace that of WASP-121. $\zeta$ Dor is a F7V star whose spectral type is similar to that of WASP-121 (F6V). The Ly$\alpha$ profile of $\zeta$ Dor is referenced from \citet{2005ApJS..159..118W}. Based on this, we constructed a double Gaussian profile with the full width at half maximum (FWHM) = 0.7 $\rm\AA$, and the two centers being 1215.5 $\rm\AA$ and 1215.9 $\rm\AA$ (Ly$\alpha_1$, see the solid black line in Figure.1(a)). The Ly$\alpha$ integrated flux is about 71900 erg cm$^{-2}$ s$^{-1}$.  The $\bar{J}_{Ly\alpha}$ is then calculated by the Ly$\alpha$ resonant scattering method of \citet{2017ApJ...851..150H}.
Since the stellar Ly$\alpha$ profile is an important physical input in calculating H(2p) population but can not be measured precisely, we investigated another Ly${\alpha}$ profile (Ly$\alpha_2$, see the dashed red line in Figure.1(a)) with the same integrated flux and two centers but with a different FWHM = 0.45 $\rm\AA$. The Ly$\alpha$ intensity is higher around the line center of Ly$\alpha_2$ profile, which leads to higher Voigt line profile weighted mean intensity (see Figure.1(b) for $\bar{J}_{Ly\alpha}$ of Ly$\alpha_1$ and Ly$\alpha_2$ ). The H$\alpha$ absorption (see Figure.1(c); for more details, see Method below) is slightly deeper for Ly$\alpha_2$ case compared to that of Ly$\alpha_1$.
The difference of absorption at the H$\alpha$ line center is less than 0.2\% and it is indistinguishable at the line wings. Our results show that the FWHM of Ly$\alpha$ input profile in the range of 0.45 to 0.7 $\rm\AA$  has minor influence on the final results. Therefore, we use Ly$\alpha_1$ in our models.

Finally, the ionization of H(2s) and H(2p) is an important process which is caused by the photons in the wavelength range of 912-3646 $\rm\AA$. The spectrum in this wavelength range is taken from the stellar atmosphere model of \citet{2003IAUS..210P.A20C}. The photoionization cross-sections of H(2s) and H(2p) are cited from TOPbase of The Opacity Project \citep{1992RMxAA..23..107C,1993A&A...275L...5C}.

\subsection{H$\alpha$ radiative transfer}

After we obtained the hydrogen populations, we simulated the H$\alpha$ radiative transfer as the stellar H$\alpha$ line travels along the ray path in the planetary atmosphere during transit \citep{2019ApJ...880...90Y}. The transmission spectrum is defined by Equation (3) as a function of wavelength,
\begin{equation}
TS(\lambda) = \frac{F_{IT}}{F_{OT}}(\lambda)-1.0
\end{equation}
and the excess absorption (not including the planet itself) is ``$-TS(\lambda)$", where $F_{IT}$ and $F_{OT}$ are the in- and out-of transit flux.
H$\alpha$ absorption is a bound-bound transition whose line center is at 6562.8 $\rm\AA$ (in air). The calculation of cross-section of H$\alpha$ absorption is similar to that of Ly$\alpha$ \citep{2019ApJ...880...90Y}, and the oscillator strength is 0.64108 at 6562.8 $\rm\AA$ taken from NIST Atomic Spectra
Database (https://www.nist.gov/pml/atomic-spectra-database).

\section{Results} \label{sec:result}
\subsection{The fiducial model}\label{sec:results_1}
We define the model with the fiducial inputs as our fiducial model (OFM). The mass loss rate of WASP-121b is 1.28$\times$10$^{12}$g/s for the fiducial model.
The number density of hydrogen atoms is obtained from the hydrodynamic simulation. The atmospheric structures are showed in Figure.2(a-c). As we can see, the highest temperature is higher than 10000 K. The velocity becomes supersonic when the altitude is larger than 1.7 R$_p$, and reaches 100 km/s beyond 7 R$_p$.
By solving a detailed equation of statistical equilibrium, we obtained the number densities of H(2s) and H(2p) (see Figure.2(a)). The number density of H(2s) plus H(2p) is about a few 10$^{-7}$ times H(1s). The ratio of H(2p) to H(2s) changes significantly with the increase of r/R$_p$, mainly due to the large photoionization of the n=2 state hydrogen, which especially affects H(2s).
The optical depth of H$\alpha$ line center is shown in Figure.2(d), which is larger than unity when r/R$_p$ is less than 1.1.

We compared the simulated transmission spectrum with the observation of \citet{2020arXiv201101245B} (B20) \footnote{Note the data of B20 used in this paper was extracted from Figure.9 of \citet{2020arXiv201101245B} by ``WebPlotDigitizer", a tool for extracting the data points (https://automeris.io/WebPlotDigitizer/index.html).}.
In their work, they reported the observations in 1-UT and 4-UT mode, the observation time of which is 06 Jan 2019 and 30 Nov 2018.
The H$\alpha$ absorption depth was about 1.4\% and 1.7\% for 1-UT and 4-UT mode, with a blue shift of 4.64 and 3.9 km/s, respectively. B20S is obtained by shifting the observed transmission spectrum of \citet{2020arXiv201101245B} towards the red side by the corresponding blue shift for 1-UT and 4-UT. 

To investigate the contribution of different atmospheric regions to the H$\alpha$ absorption, we calculated the H$\alpha$ absorption for different altitudes of the atmosphere until the altitude reaches 7.6 R$_p$.
Figure.3(a) shows the results of the fiducial model, in which the gray dots with errorbars are B20S. The lightgray and darkgray points are for 1-UT and 4-UT, respectively.
The different lines represent the absorption of H$\alpha$ produced in different atmospheric altitudes. Our simulations show that different altitudes of the atmosphere contribute differently to the final transmission spectrum of H$\alpha$. The increase of the altitude leads to a deeper absorption. From 3 R$_p$ to 4 R$_p$, the absorption at H$\alpha$ line center only increases by 0.05\%.
The H$\alpha$ absorption of WASP-121b by the atmosphere above 4 R$_p$ is negligible because the H(2) are sparse.
In addition, we compared the equivalent width (EW) of the  model  transmission spectrum with that of the observations in the passbands 0.75, 1.0, 1.5, 2.0, 2.5, 3.0 and 3.6 $\rm\AA$. We found that the EW will decrease with the increase of the passbands when the passbands are larger than 1.5 $\rm\AA$ for 1-UT, due to some minus values of the absorption at the line wings. Besides, we also found that the EW calculated by our model is lower than that of the 4-UT if the passband is greater than 1.5 $\rm\AA$. Thus, we only  analysed the EW in passbands 0.75, 1.0 and 1.5 $\rm\AA$. Figure.3 (b), (c) and (d) show the equivalent width calculated in the three passbands as a function of the atmospheric altitude, respectively. 
The black line represents the fiducial model, in which the sonic point is marked by the purple cross. The orange and cyan horizontal solid lines represent the mean EW of 1-UT and 4-UT, respectively. The corresponding dashed lines are the upper (+ 1$\sigma$) and lower (- 1$\sigma$) limit of the observations. One can see that the EW of the fiducial model increases with the increase of the atmospheric altitude. For 1-UT, an atmosphere higher than 1.7 Rp can fit the observation in the three passbands. For 4-UT, a supersonic atmosphere beyond the Roche lobe (1.86 R$_p$) is required to fit the EW in 0.75 and 1.0 $\rm\AA$. For example, an atmosphere at least up to 6 R$_p$ can fit the lower limit of the observation in passband 1.0 $\rm\AA$. The EW of the model in passband 1.5 $\rm\AA$, however, can not reach the lower limit of the 4-UT observation, because our model can not reproduce the relatively high absorption at the H$\alpha$ line wings.

The results above show that the absorption of H$\alpha$ close to the line center can be well fitted by our model and the contribution of supersonic regions can not be neglected. 
However, the strong absorption in the wings of H$\alpha$ for 4-UT should be investigated in the future. Furthermore, the 1D simulation can not reproduce the blue-shift found by B20.
A blue- or red- shift has also been found in other absorption lines in exoplanetary atmosphere \citep{2018A&A...616A.151C,2018Sci...362.1384A,2020MNRAS.493.2215G,2020A&A...635A.205B,2020MNRAS.494..363C}, which can be led by the atmospheric winds or the circulations from the day-side to night-side \citep{2013ApJ...766..102}.

\subsection{XUV integrated flux}\label{sec:results_1}
Since the XUV integrated flux was calculated by the age-luminosity relation of \cite{2011A&A...532A...6S}, there could be some uncertainties due to the uncertainty of the stellar age \citep{2016MNRAS.458.4025D}.
Here we investigated to what extent the $F_{XUV}$ will influence the H$\alpha$ transmission spectrum. We adopted 0.5, 0.75, 1.5, 2.0, 3.0 and 4.0 times the value of fiducial $F_{XUV}$  ($F_{0}$), and the mass loss rates are 8.20$\times$10$^{11}$g/s, 1.07$\times$10$^{12}$g/s, 1.58$\times$10$^{12}$g/s, 1.84$\times$10$^{12}$g/s, 2.29$\times$10$^{12}$g/s and 2.68$\times$10$^{12}$g/s accordingly. The energy-limited theory \citep{2003ApJ...598L.121L, 2007A&A...472..329E} proposes that the mass loss rates of the atmosphere are proportional to the $F_{XUV}$. However, our results found that they do not increase linearly with the increase of the $F_{XUV}$ because the escape of Ly$\alpha$ photons takes away a portion of the heat from the atmosphere. The H$\alpha$ absorption increases moderately with $F_{XUV}$ (see Figure.4(a), the transmission spectra are calculated within 7.6 R$_p$). For instance, an increase of a factor of 8 in $F_{XUV}$ only increases the absorption at line center by $\sim$ 0.5\%.
Two reasons can explain this phenomenon.
For higher $F_{XUV}$, the corresponding temperature of the atmosphere becomes higher at the bottom of the atmosphere but will drop dramatically with the increase of atmospheric altitude. The high temperature occurs in the relatively high pressure and is close to the bottom of the atmosphere, so that the Ly$\alpha$ photons will spend a longer time to be scattered out of the atmosphere. Therefore, the $\bar{J}_{Ly\alpha}$ will be more intense and then will excite more hydrogen atoms into the n=2 state. However, the atoms can be ionized easily owing to higher $F_{XUV}$. The combined effect is to increase the H$\alpha$ absorption slightly.
It is clear from Figure.4(a) that a higher $F_{XUV}$ is needed to fit the 4-UT data, while a relatively lower $F_{XUV}$ is preferable for the data of 1-UT. This can be verified by Figure.4(b), which shows the $\chi ^2$ as a function of $F_{XUV}$. The $\chi ^2$ is calculated in passband 1.5 and 3.6 $\rm\AA$ for 1-UT and 4-UT, respectively.
The minimum $\chi ^2$ for 1-UT and 4-UT occurs at $F_{XUV}$ = F$_0$* 0.5 and $F_{XUV}$ = F$_0$* 1.5, respectively. \cite{2020arXiv201101245B} reported that log$R^{\prime}_{HK}$  = -4.87 $\pm$ 0.01 and log$R^{\prime}_{HK}$  = -4.81 $\pm$ 0.01 for the 1-UT and 4-UT transits, respectively, and proposed that the star was more active at 4-UT transit. Therefore, the different $F_{XUV}$ levels for 1-UT and 4-UT may reflect the different stellar activities of WASP-121 in the two observations. A higher $F_{XUV}$ of 4-UT could probably attributes to its higher activity level compared to that of 1-UT.

We also simulated the H$\alpha$ transmission spectra for the cases of $F_{XUV}$ = F$_0$* 0.5 and $F_{XUV}$ = F$_0$* 1.5 as a function of different altitudes. Figure.5(a) shows the case of $F_{XUV}$ = F$_0$* 0.5. The absorption caused by the atmosphere below the sonic point (1.7 R$_p$) is shallower compared with 1-UT of B20S. Figure.5(b) shows the equivalent width as a function of different altitudes. It shows when WASP-121b receives 0.5 times the fiducial XUV irradiation of the host star, the EW produced by the atmosphere lower than 2.5 R$_p$ can not reach the lower limit of the observation. Above 2.5 R$_p$, the more the atmosphere expands, the better the fit to the observation of 1-UT. Figure.5(c) and (d) are the same to Figure.5(a) and (b), respectively, but for the case of $F_{XUV}$ = F$_0$* 1.5 and in comparison with 4-UT of B20S. The EW of the model can not reach the lower limit of the EW of 4-UT, because our model can not reproduce the relatively high absorption at the H$\alpha$ line wings.

\subsection{XUV SEDs}\label{sec:results_1}
XUV SEDs can also influence the photoionization process in the atmosphere\citep{2016ApJ...818..107}.
According to \cite{2012MNRAS.425.2931O}, X-ray can solely drive hydrodynamic escape of planetary atmosphere. Here we study the effect of different SEDs on the transmission spectra by introducing a modified spectral index $\beta_{m}$, defined as F(1-100 $\rm\AA$)/F(1-912 $\rm\AA$), where F(1-100 $\rm\AA$) is the integrated flux in the band of X-ray and F(1-912 $\rm\AA$) is the integrated flux of the whole XUV band. For the fiducial model, the value is 0.1475. We tested the cases of 0.03 (almost no X-ray but all EUV), 0.103, 0.221, and 0.5 (half X-ray and half EUV, which could be not real according to the evolution of XUV radiation of late-type stars, and it is for model experiment), while the XUV integrated flux $F_{XUV}$ = $F_0$. Figure.4(c) shows the transmission spectra (calculated within 7.6 R$_p$). One can see that a larger X-ray proportion will lead to deeper H$\alpha$ absorption. The main reason is that more hydrogen atoms will be retained instead of being ionized, due to the lower photoionization cross section (inversely proportional to the cube of the frequency) in X-ray band in comparison with that in EUV. To compare with the observations, Figure.4(d) shows the $\chi ^2$ as a function of $\beta_{m}$. For 1-UT, the minimum $\chi ^2$ appears at $\beta_{m}$= 0.003. For 4-UT, the $\chi ^2$ decreases with the increase of $\beta_{m}$. A higher $\beta_{m}$ is required to fit the absorption at H$\alpha$ line wings for 4-UT. However, the variation of $\chi ^2$ is less than 1 for $\beta_{m}$ from 0.103 to 0.5. In addition, for the case of $\beta_{m}$= 0.5 the absorption at H$\alpha$ line center exceeds the upper limit (+ 1$\sigma$) of the observation. Thus, a high $\beta_{m}$ can not be applicable for 4-UT although the value of the $\chi ^2$ is smaller. Thus, we suggest that $\beta_{m}$ should be confined to a relatively low level in order to fit the observation at different times, which is consistent with the evolution of XUV radiation of late-type stars \citep{2011A&A...532A...6S}.

\section{Discussion} \label{sec:discuss}
In addition to the observation of H$\alpha$ transmission spectroscopy conducted by \citet{2020arXiv201101245B}, there is another observation made by \citet{2020MNRAS.494..363C} (C20). According to C20, the observed transmission spectrum can be fitted with a Gaussian profile, of which the full width at half maximum (FWHM) is 0.75$\rm\AA$ and the line center is at 6562.93 $\rm\AA$. C20S is the transmission spectrum obtained by shifting the Gaussian fitted spectrum towards the blue side by 5.82 km/s.
The average absorption depths (ADs) in different passbands are also shown in their work (see Table.3 of \citet{2020MNRAS.494..363C}).

For comparison with C20, we calculated the ADs of the simulated transmission spectra in different passbands as shown in Figure.6. The passbands are 0.188, 0.375, 0.75, 1.5 and 3 $\rm\AA$, which are the same to that used in C20. In their work, they used the photo noise and readout noise from the observed spectrum to calculate the weight \citep{2017A&A...608A.135C} for the mean absorption depth. Here, we evaluated the ADs of C20S for the above passbands by an equally weighted method (i.e., the spectral points are weighted equally despite different errors) and found that the results did not deviate much from C20. We also calculated the $\chi ^2$ of the  ADs in different passbands for the models of different $F_{XUV}$ and XUV SEDs with respect to that of C20.

Figure.6(a) shows the the $\chi ^2$ as a function of $F_{XUV}$ for different passbands. The minimum $\chi ^2$ appears at 0.5, 0.75 and 1.0 $F_0$ for the given passbands, indicating that the $F_{XUV}$ is not larger than the fiducial value. The $F_{XUV}$ values in the range of 0.5-1.0 times $F_0$ reflect that the stellar activity may be between that of 1-UT and 4-UT of B20.
We also simulated the H$\alpha$ transmission spectra for the cases of 0.5, 0.75 and 1.0 times F$_0$ and found that the H$\alpha$ absorption caused by the atmosphere below the sonic points for the three $F_{XUV}$ cases is not enough to match C20S, especially for the absorption at line center (see Figure.6(b)), which shows that the supersonic regions are not negligible in explaining the excess absorption of H$\alpha$ of WASP-121b.

Figure.6(c) shows the $\chi ^2$ as a function of $\beta_{m}$ for different passbands.
For the cases of 0.188, 0.375 and 1.5 $\rm\AA$, the minimum
$\chi ^2$ appears at $\beta_{m}$ = 0.103; for the cases of 0.75 and 3.0 $\rm\AA$, the model of  $\beta_{m}$= 0.03 is closest to the observation.
This is consistent with the conclusion of B20S that $\beta_{m}$ is confined to a lower level.

\section{Summary} \label{sec:summary}
In this paper, we presented the XUV driven hydrodynamic simulation including the detailed hydrogen population calculation and the process of radiative transfer to model the H$\alpha$ transmission spectrum of WASP-121b. Our models are in agreement with the observations.
We found that the supersonic regions of planetary wind contribute a prominent portion to the absorption of H$\alpha$.
We also performed a broad parameter study to evaluate the affects of the input stellar XUV integrated flux and SEDs.
Our results showed that the variations of the stellar $F_{XUV}$ can be in the range of 0.5-1.5 times the fiducial value and the different $F_{XUV}$ level inferred from the independent observations may reflect the stellar activities of the host star.
It also showed that the X-ray portion in the XUV radiation should be at a low level, which is consistent with the evolution of XUV radiation of late-type stars \citep{2011A&A...532A...6S}. The parameter study enhanced the conclusion of the fiducial model that the supersonic regions are indispensable in the interpretation of the excess absorption of H$\alpha$ for WASP-121b, which clearly expresses the requirement of a transonic hydrodynamic atmosphere. The consistence of our simulations and the observation of H$\alpha$ transmission spectrum suggested that there is an expanding hydrogen atmosphere around this planet. These findings are helpful for the future detection of the escaping planetary atmosphere around F-type stars by using the ground telescope.

\vspace{12 pt}
\textbf{Acknowledgement.} We thank the anonymous reviewers for their constructive comments to improve the manuscript. We also thank Z. W. Han and G. Chen for their helpful discussions about the observation data. The authors also acknowledge supports by the Strategic Priority Research Program of Chinese Academy of Sciences, Grant No. XDB 41000000 and from National Natural Science Foundation of China though grants 11973082 to JG.

\clearpage

\begin{figure*}
\gridline{\fig{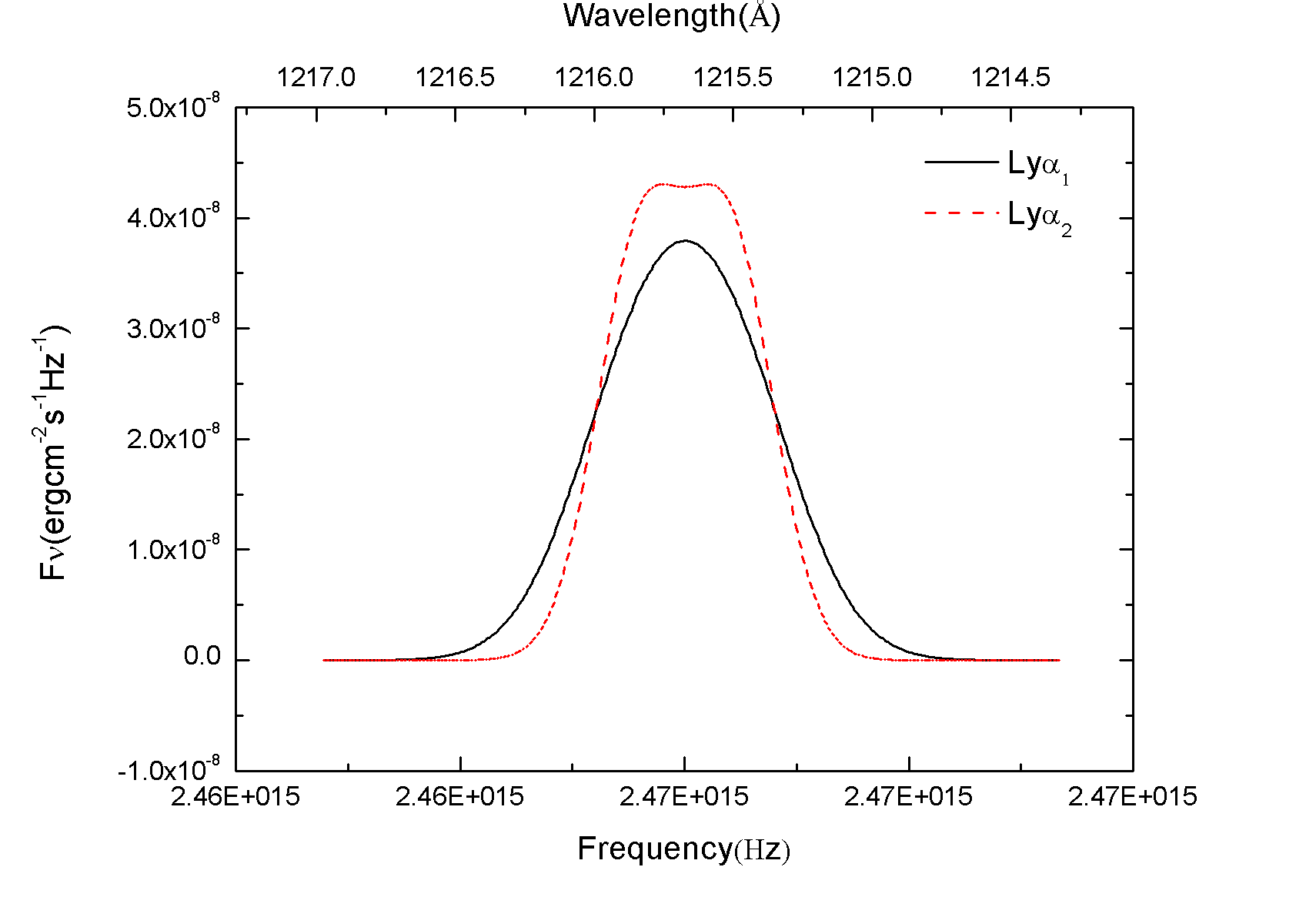}{0.6\textwidth}{(a)}}
\gridline{\fig{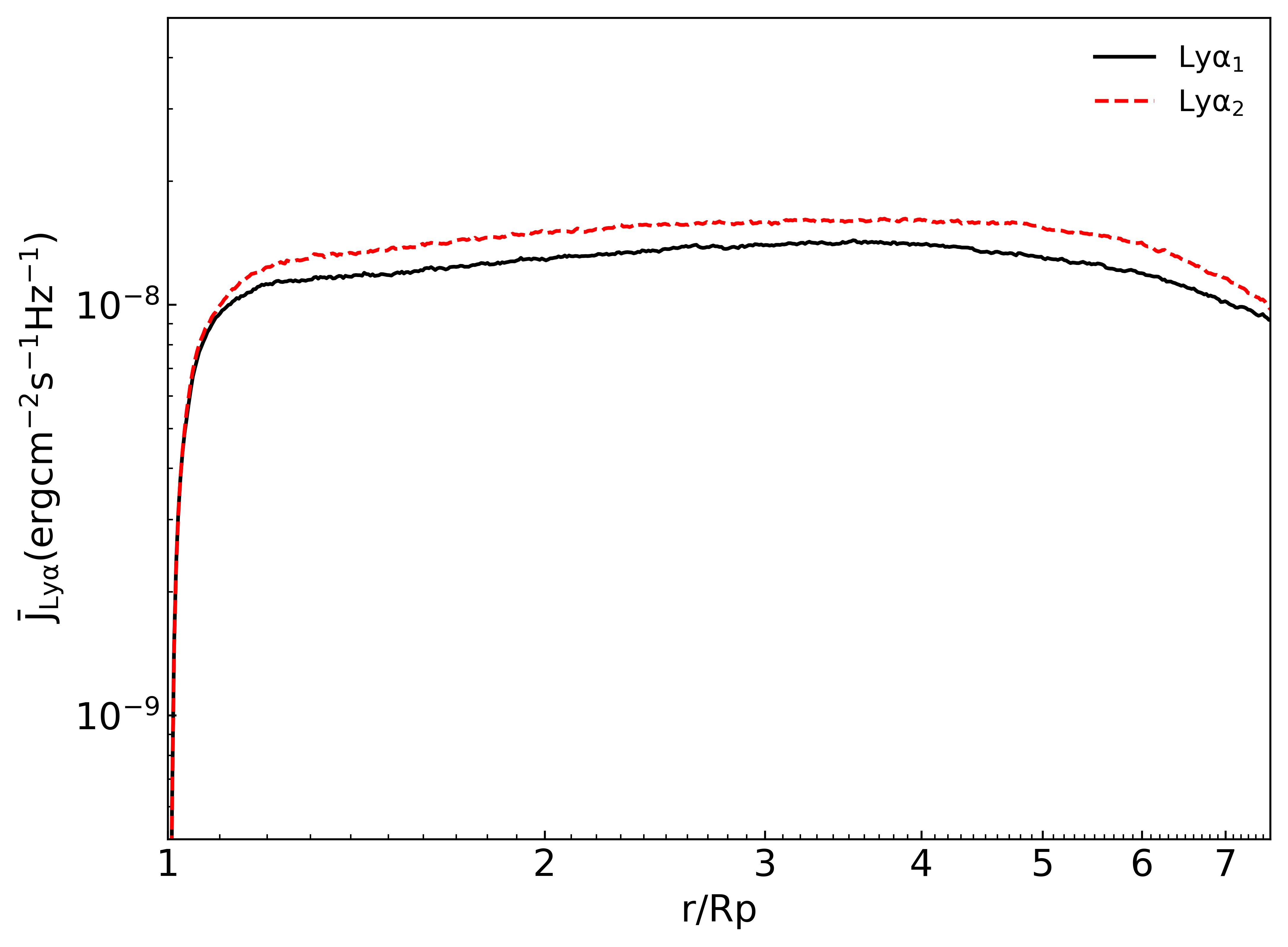}{0.5\textwidth}{(b)}}
\gridline{\fig{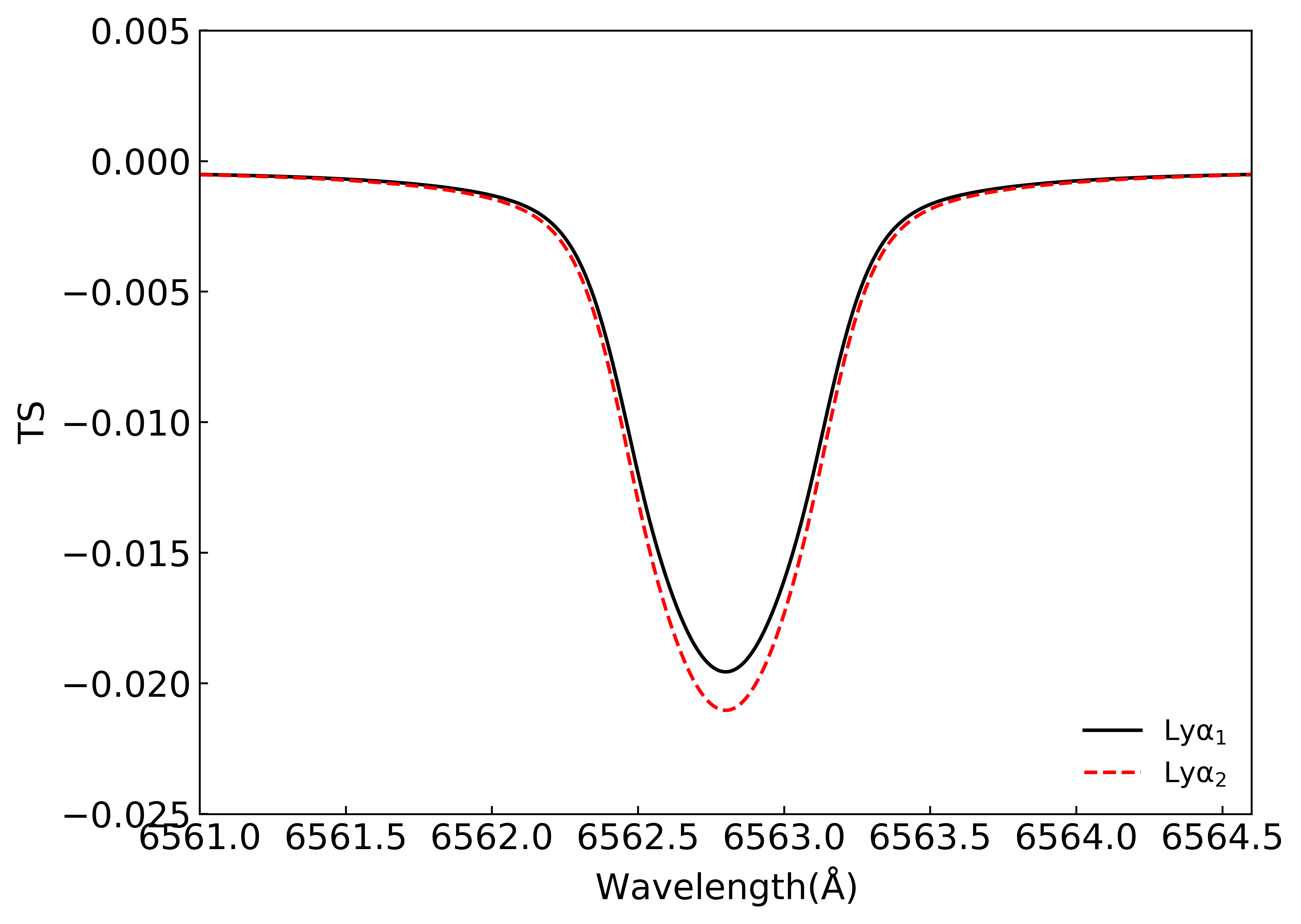}{0.5\textwidth}{(c)}}
\caption{\textbf{Properties of stellar Ly$\alpha$.} (a)The stellar Ly$\alpha$ profiles. The FWHM of Ly$\alpha_1$ and Ly$\alpha_2$ are 0.7$\rm\AA$ and 0.45$\rm\AA$, respectively. (b) The Ly$\alpha$ Voigt line profile weighted mean intensity for Ly$\alpha_1$ and Ly$\alpha_2$. (c) The H$\alpha$ transmission spectrum for Ly$\alpha_1$ and Ly$\alpha_2$.}\label{sup_Lya}
\end{figure*}

\begin{figure*}
\gridline{\fig{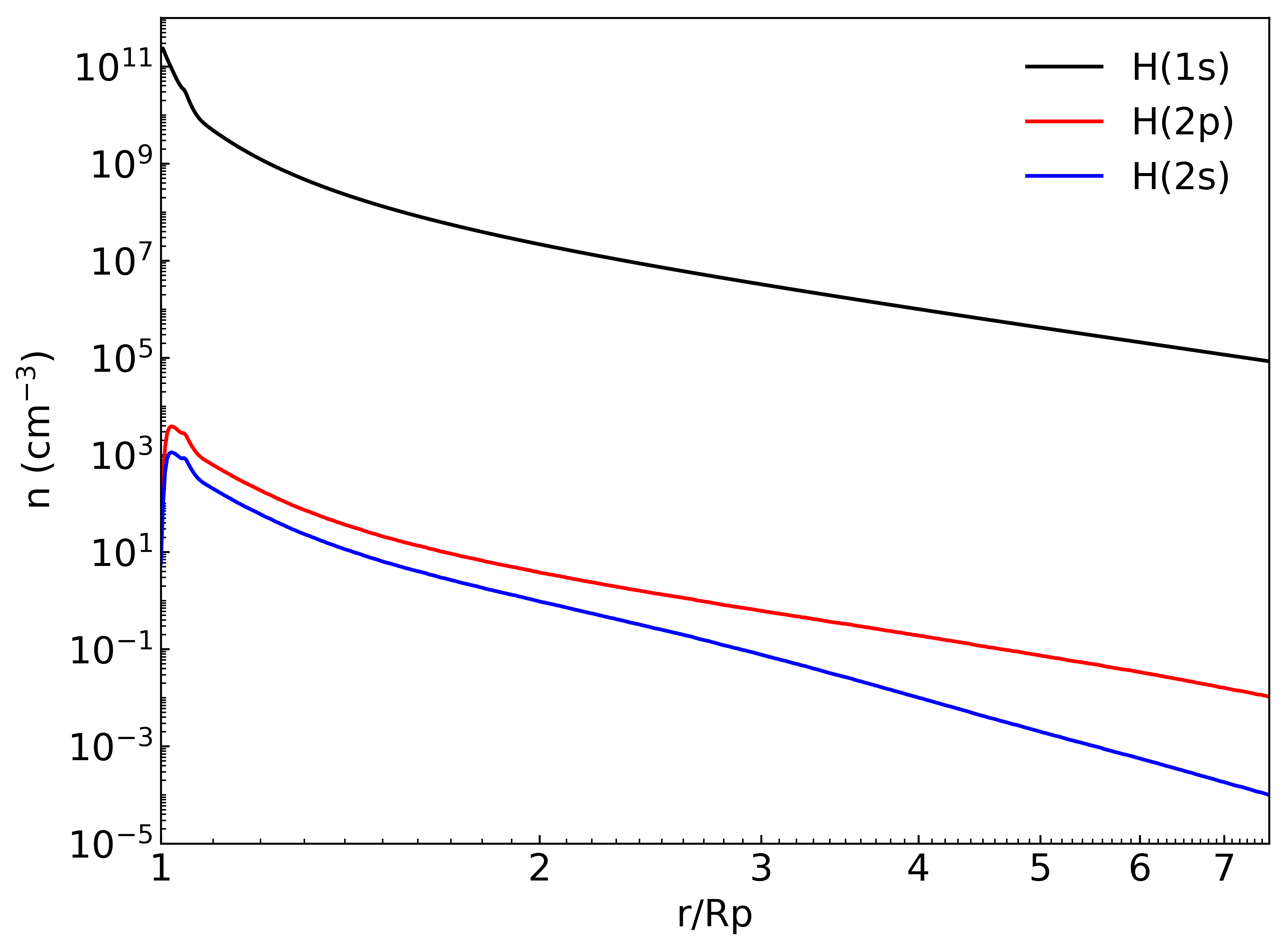}{0.5\textwidth}{(a)}
          \fig{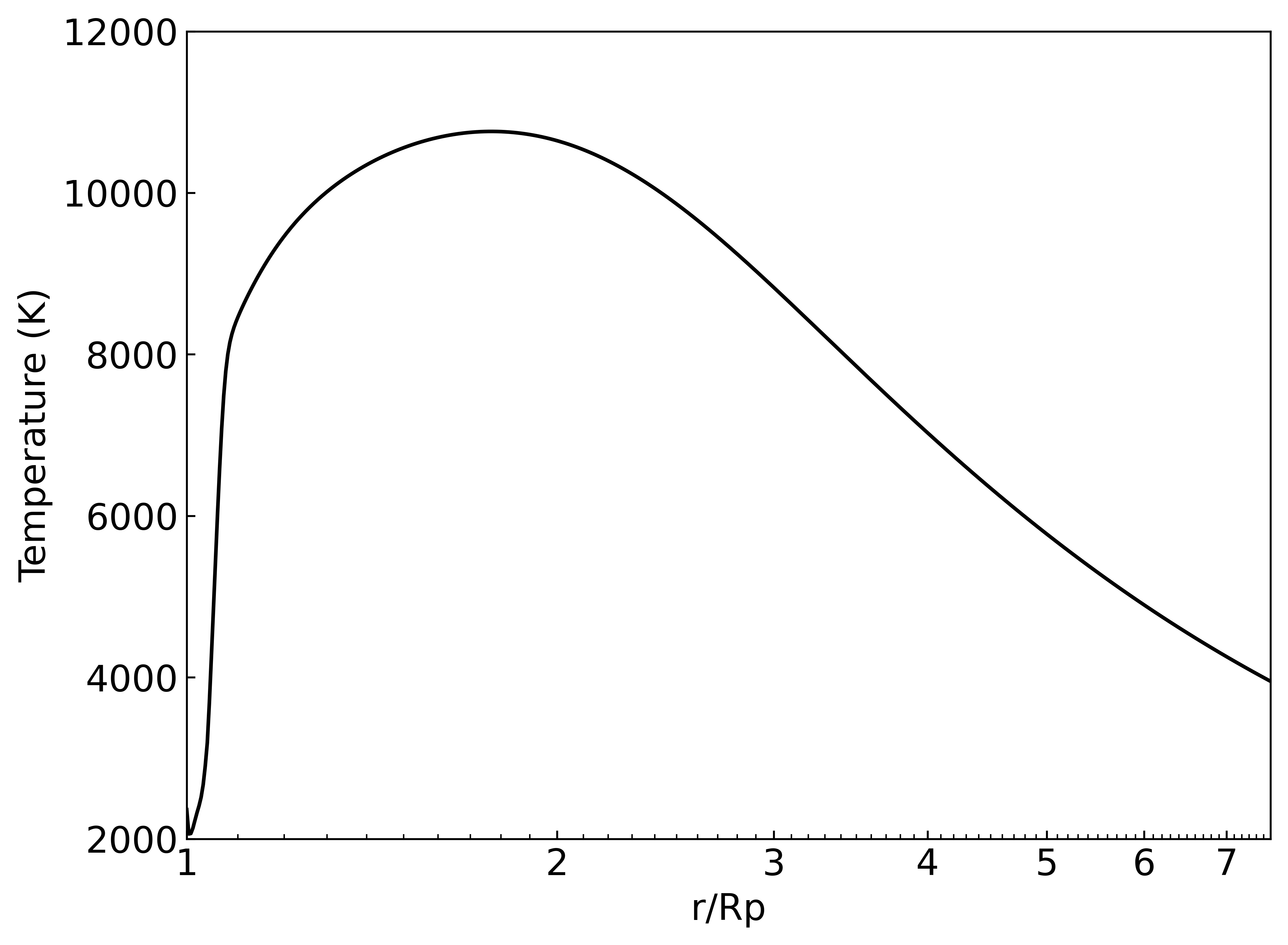}{0.5\textwidth}{(b)}
          }
\gridline{\fig{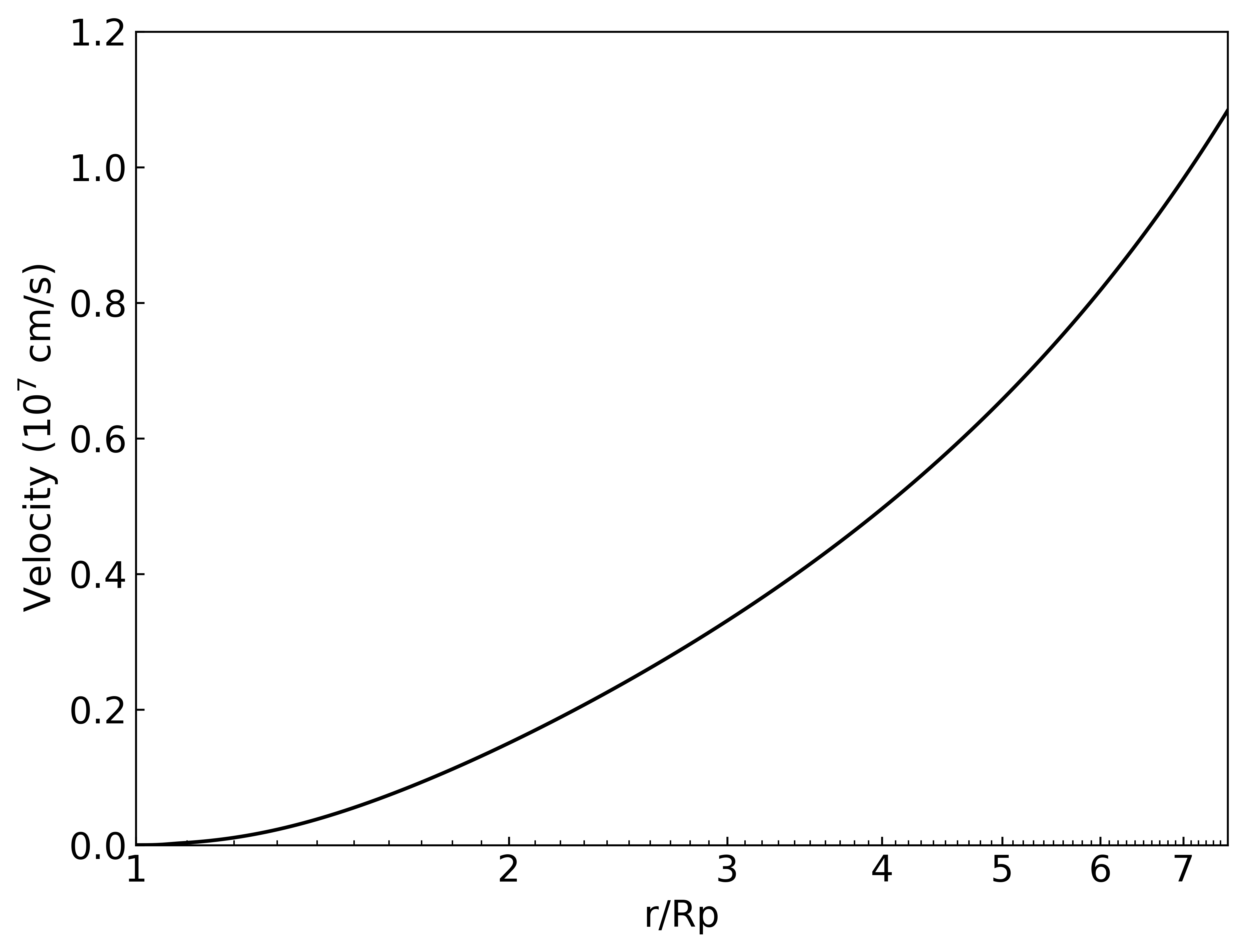}{0.5\textwidth}{(c)}
          \fig{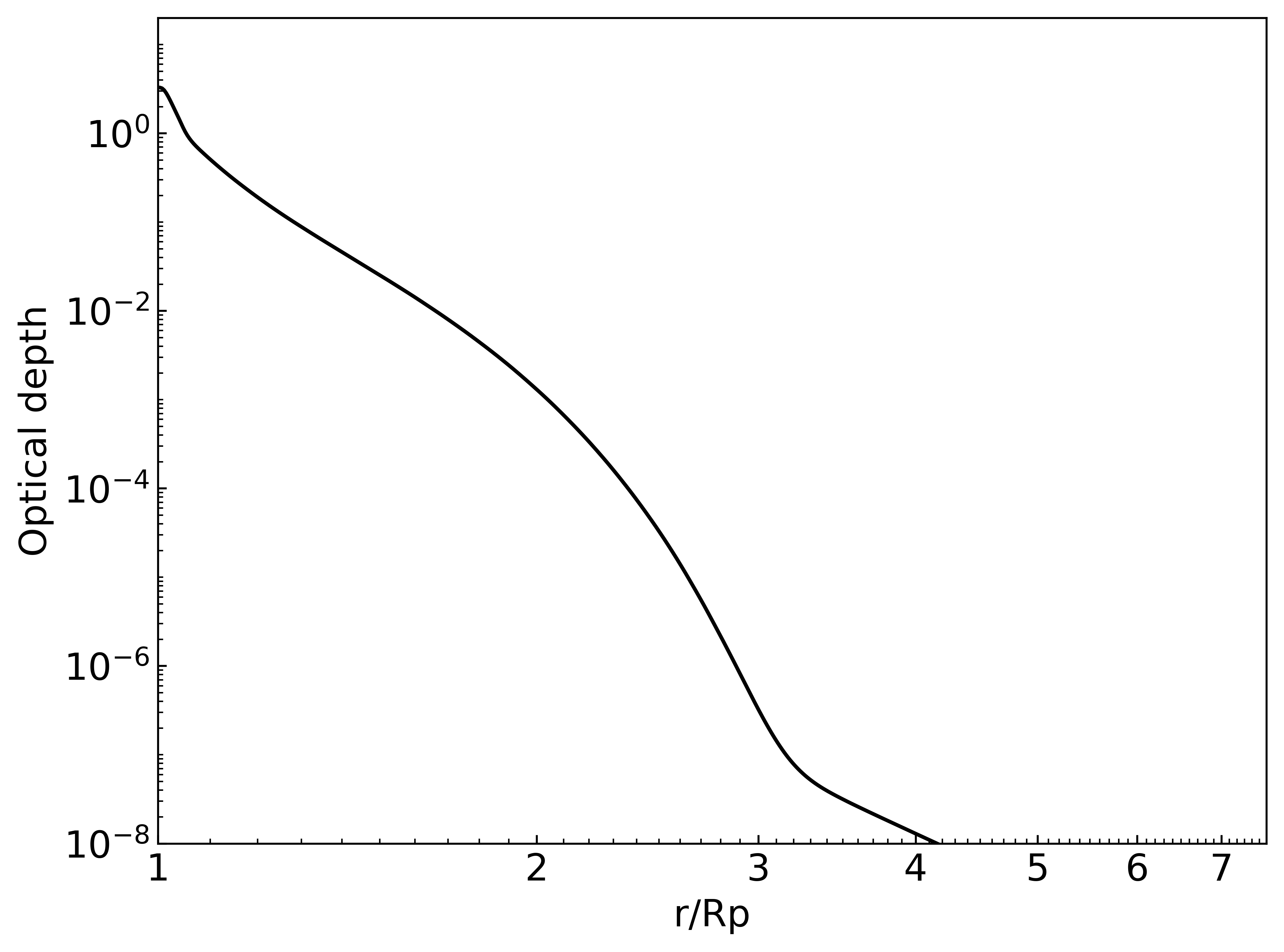}{0.5\textwidth}{(d)}
          }
\caption{\textbf{Atmospheric structures.} This structure is plotted as a function of r/Rp, r is the distance from the planetary center and Rp is the planetary radius.
(a) The number density for H(1s), H(2p) and H(2s). (b) The atmospheric temperature. (c) The particles' velocity. (d) Optical depth of H$\alpha$ line center as a function of impact parameter.} \label{fidu_atm}
\end{figure*}

\begin{figure*}
\gridline{\fig{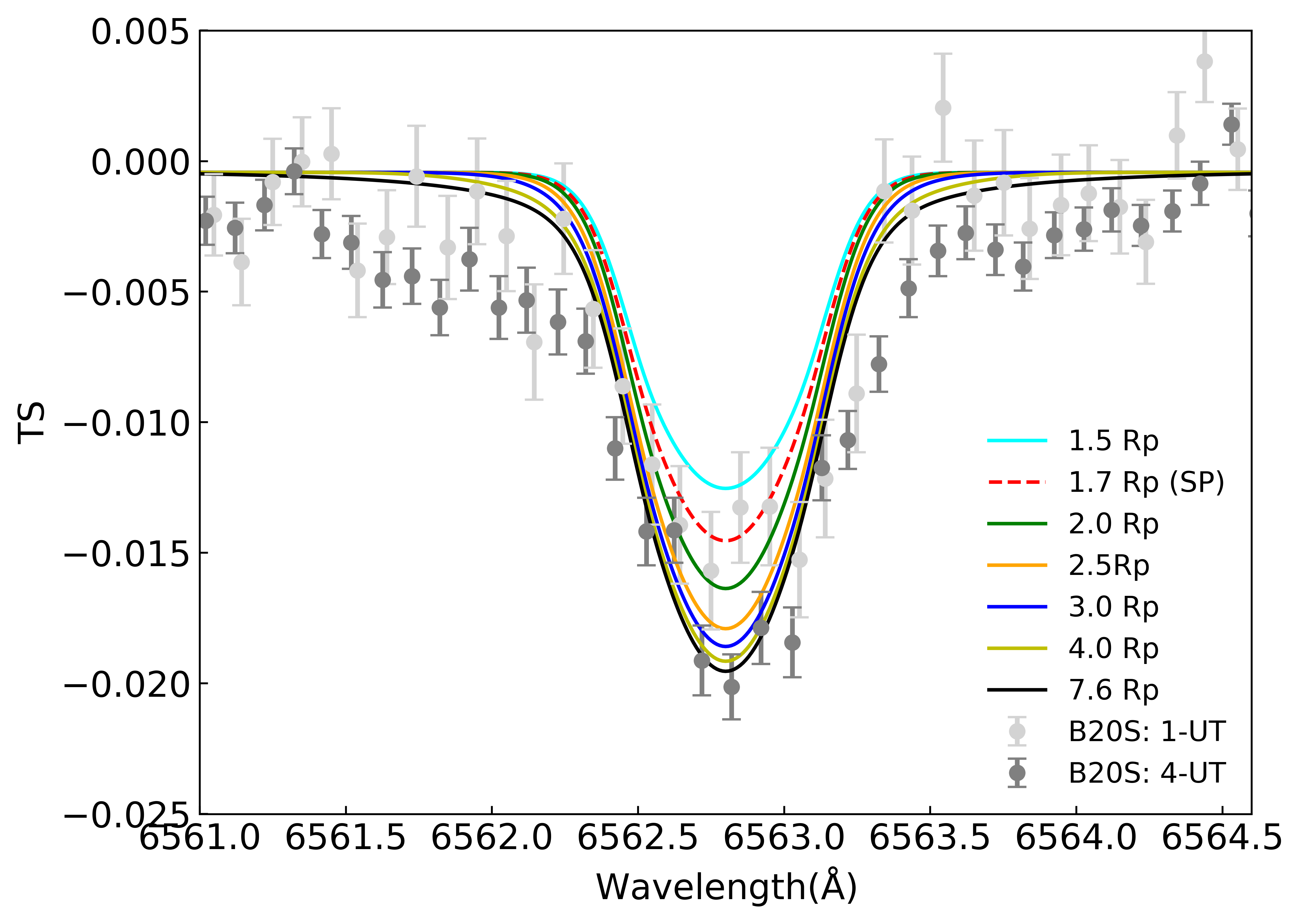}{0.5\textwidth}{(a)}
          \fig{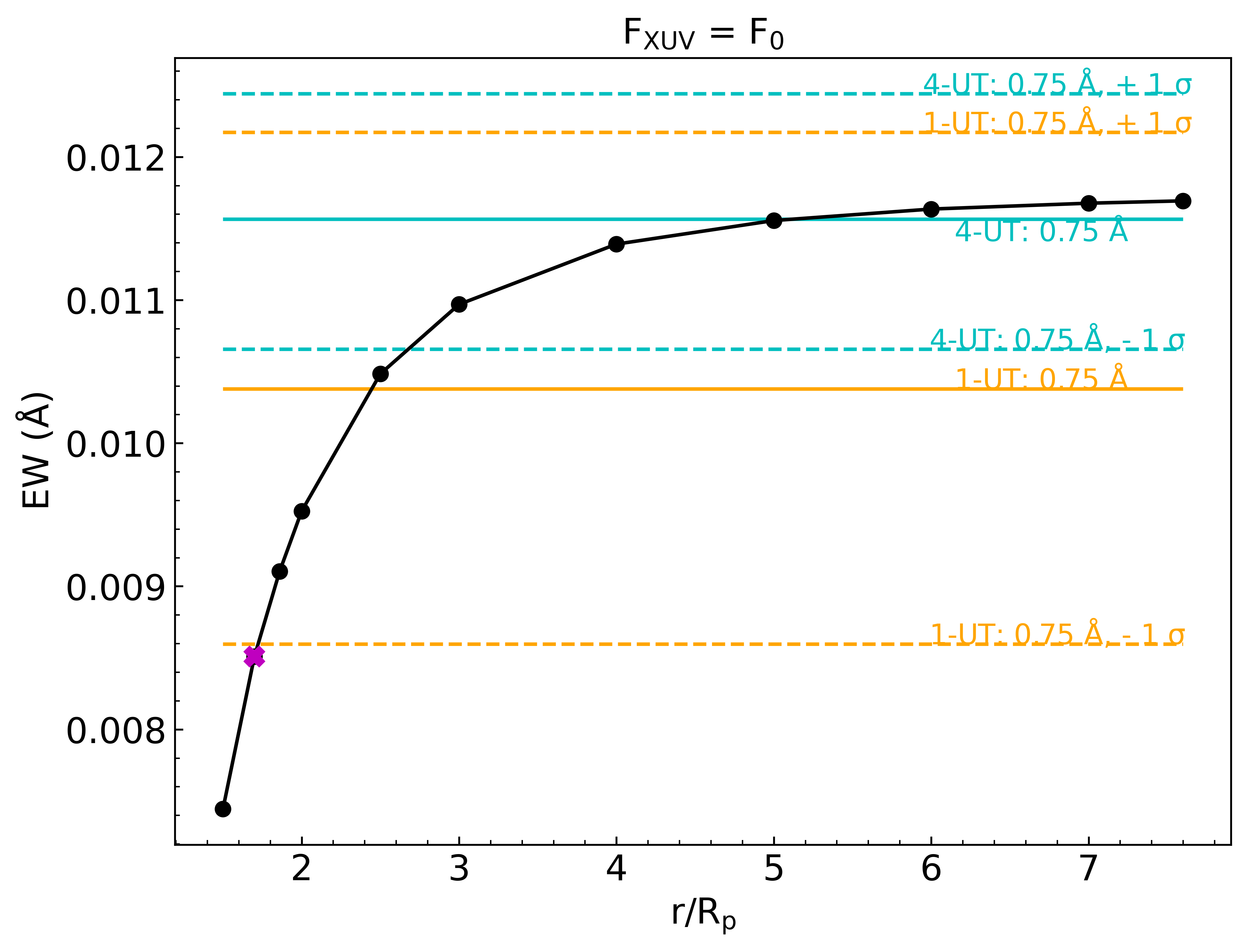}{0.5\textwidth}{(b)}
          }
\gridline{\fig{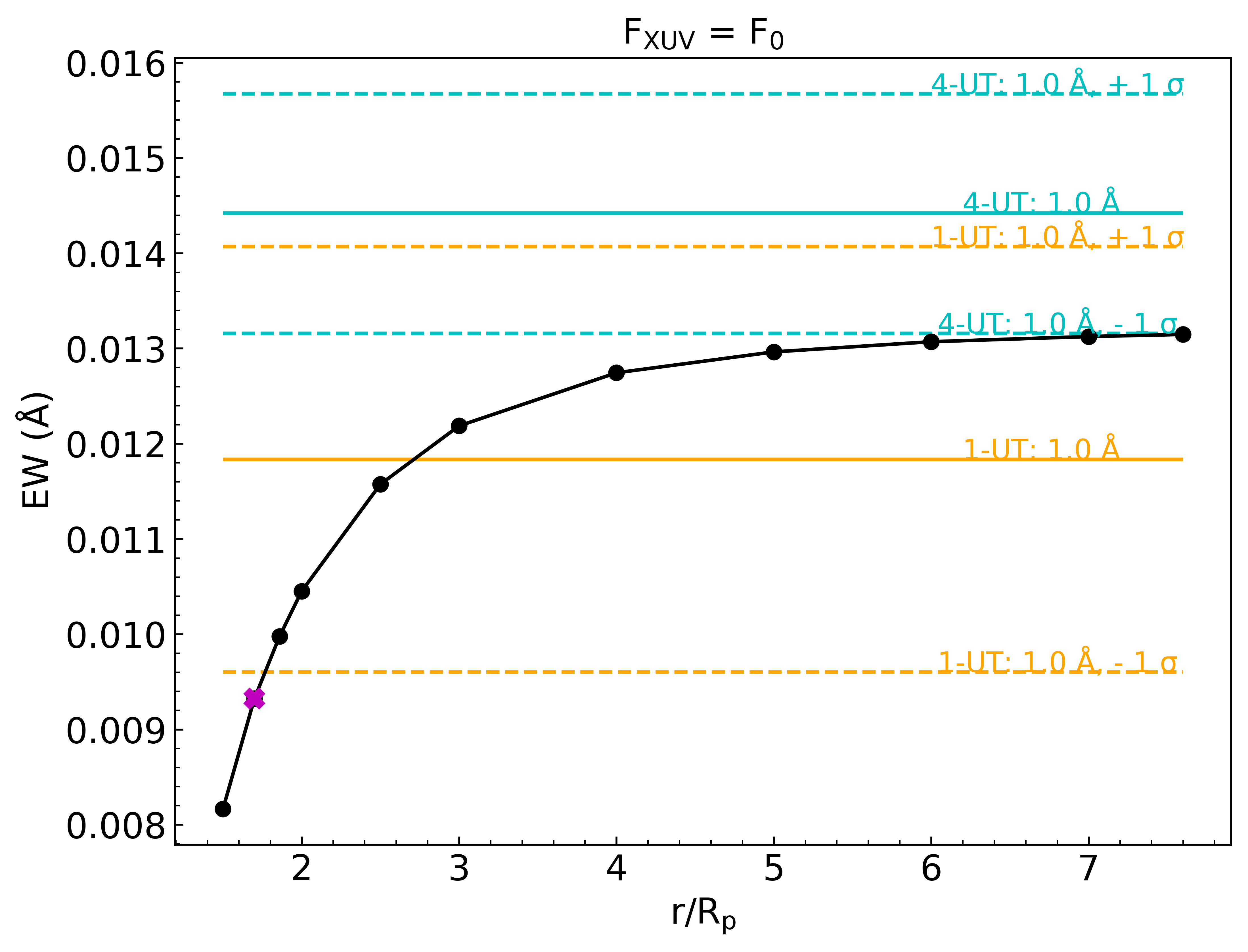}{0.5\textwidth}{(c)}
          \fig{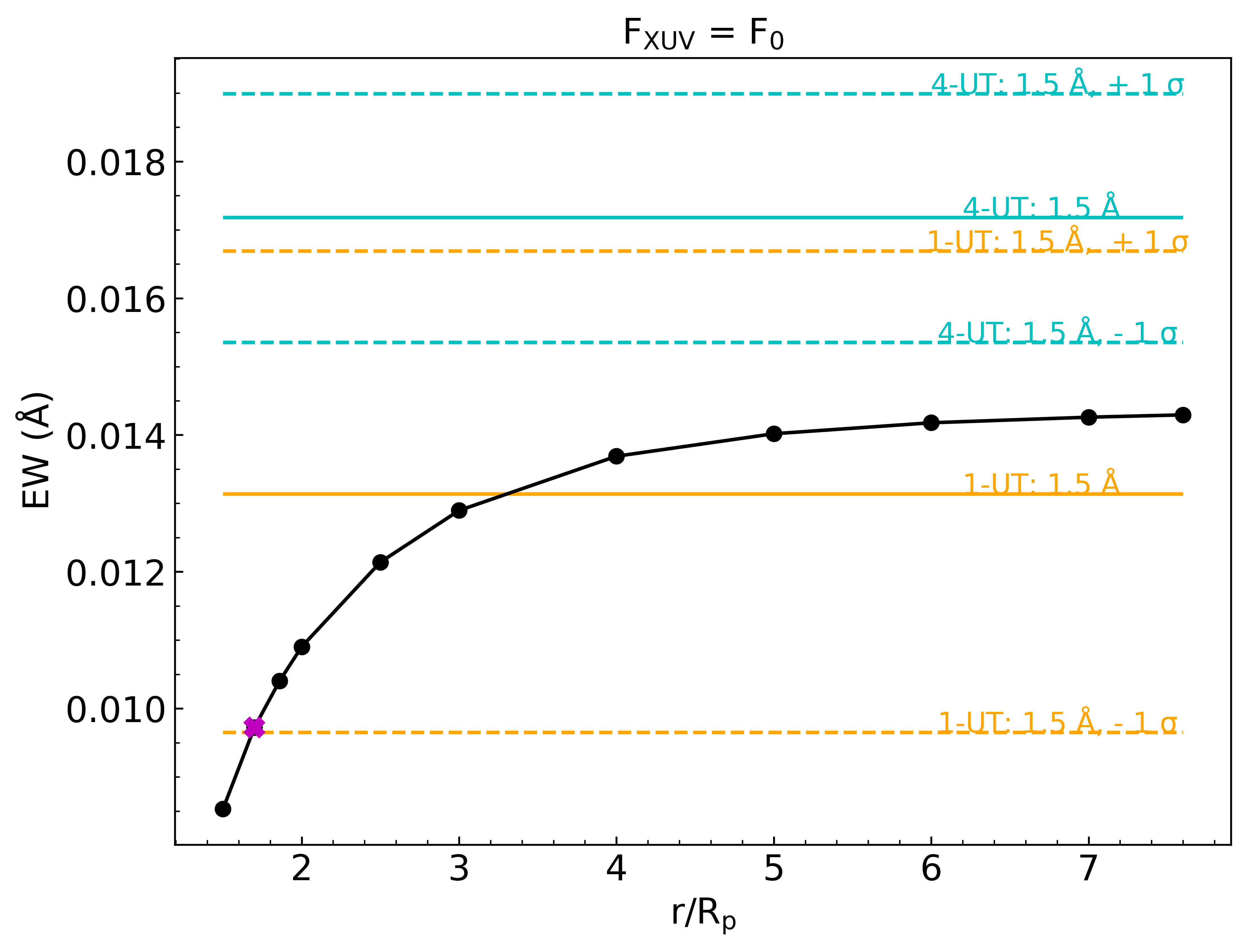}{0.5\textwidth}{(d)}
          }
\caption{\textbf{H$\alpha$ transmission spectrum and the equivalent width (EW).} (a) H$\alpha$ transmission spectrum. The gray dots with errors are the H$\alpha$ transmission spectrum of B20S, with the lightgray for 1-UT and darkgray for 4-UT.
The red dashed line represents the absorption within the sonic point (SP). Other lines represent the absorption within the labeled atmosphere regions. (b) The equivalent width calculated in passband 0.75 $\rm\AA$ as a function of atmospheric altitude. The black line represents the EW of the fiducial model, in which the sonic point is marked by the purple cross. The orange and cyan horizontal solid lines represent the mean EW of 1-UT and 4-UT, respectively. The corresponding dashed lines are + 1 $\sigma$ and - 1 $\sigma$ of the mean EW. (c) and (d) are the same as (a), but for the passbands 1.0 and 1.5 $\rm\AA$, respectively.} \label{fidu_trans}
\end{figure*}

\begin{figure*}
\gridline{\fig{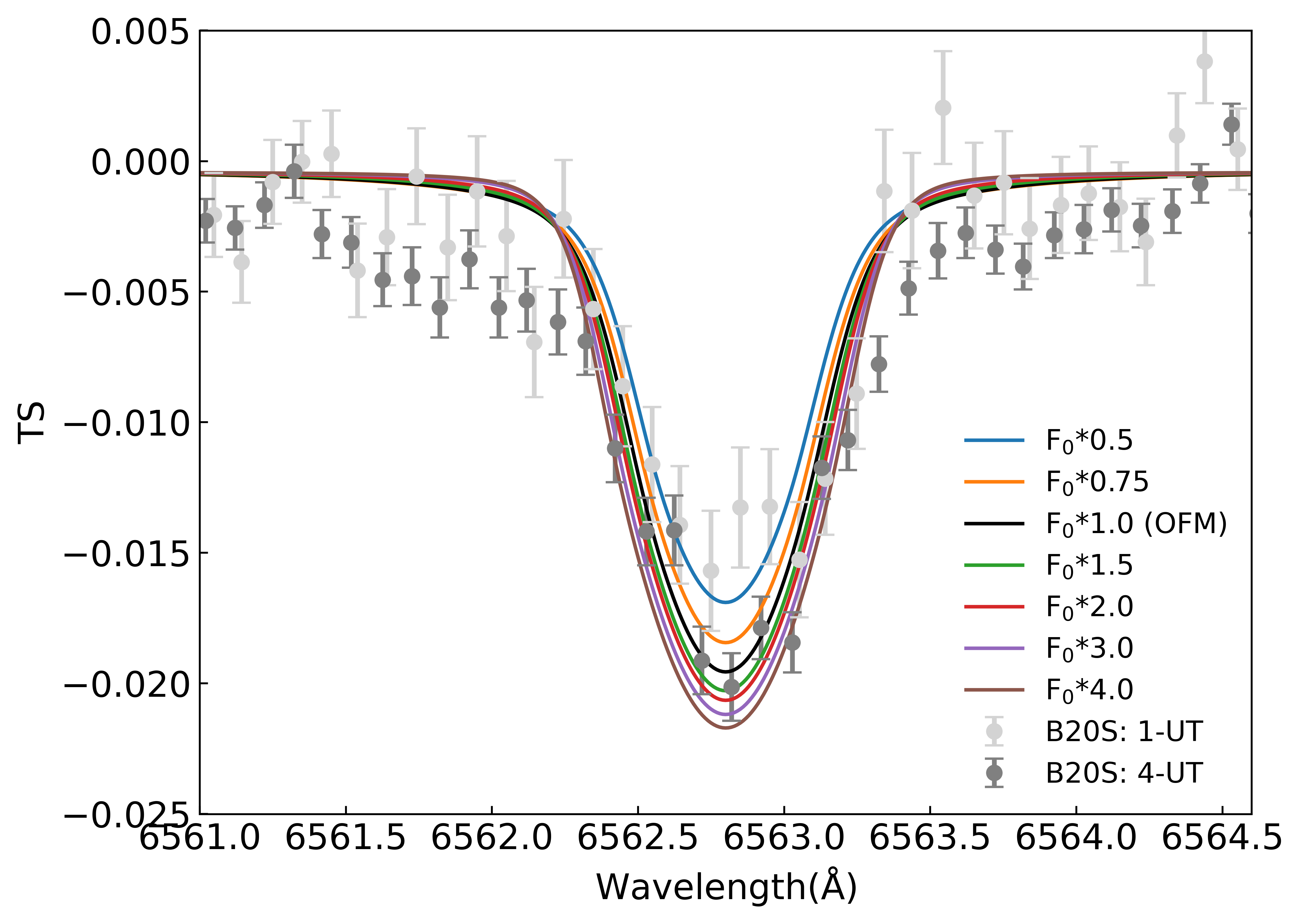}{0.5\textwidth}{(a)}
          \fig{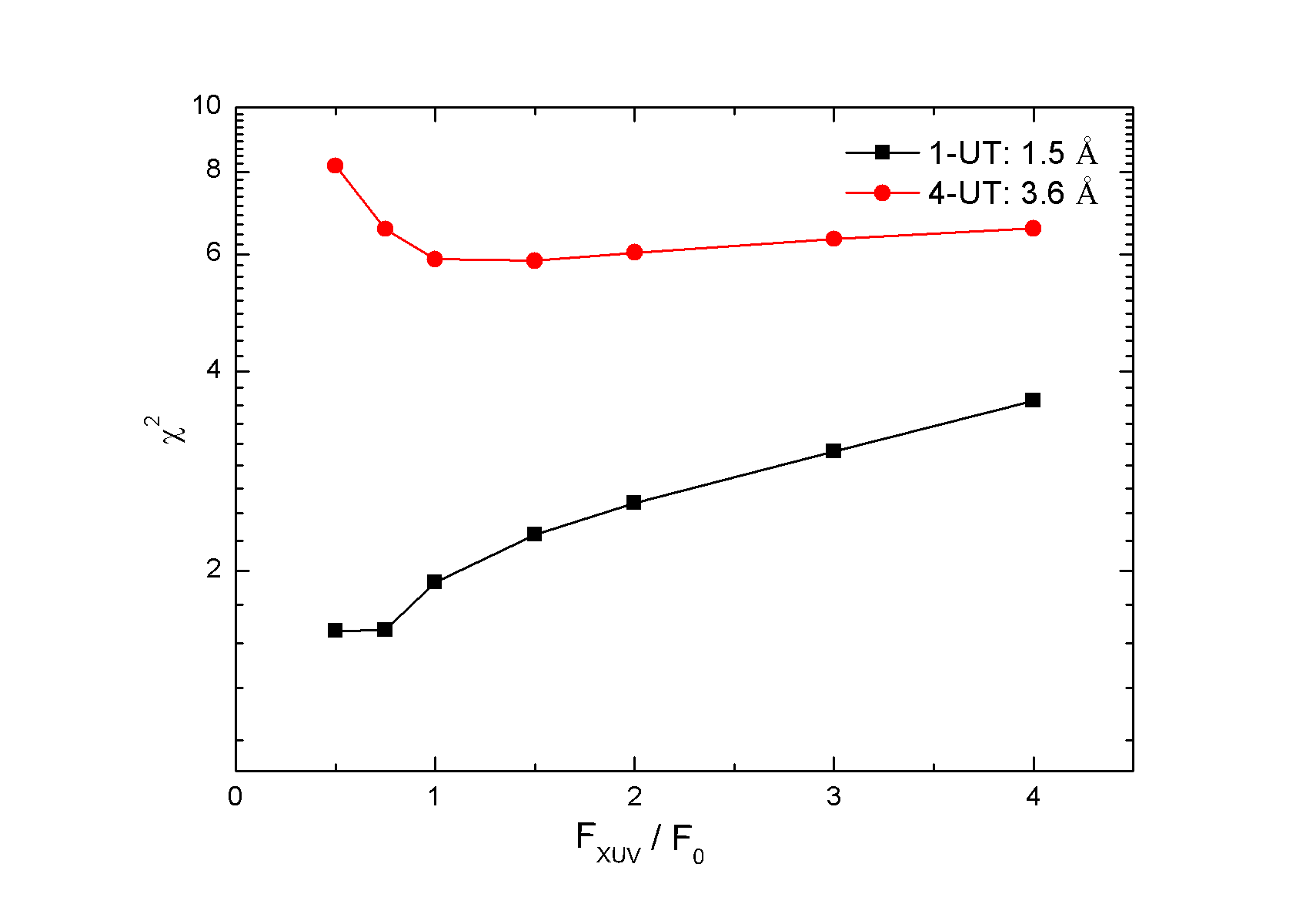}{0.6\textwidth}{(b)}
         }
\gridline{\fig{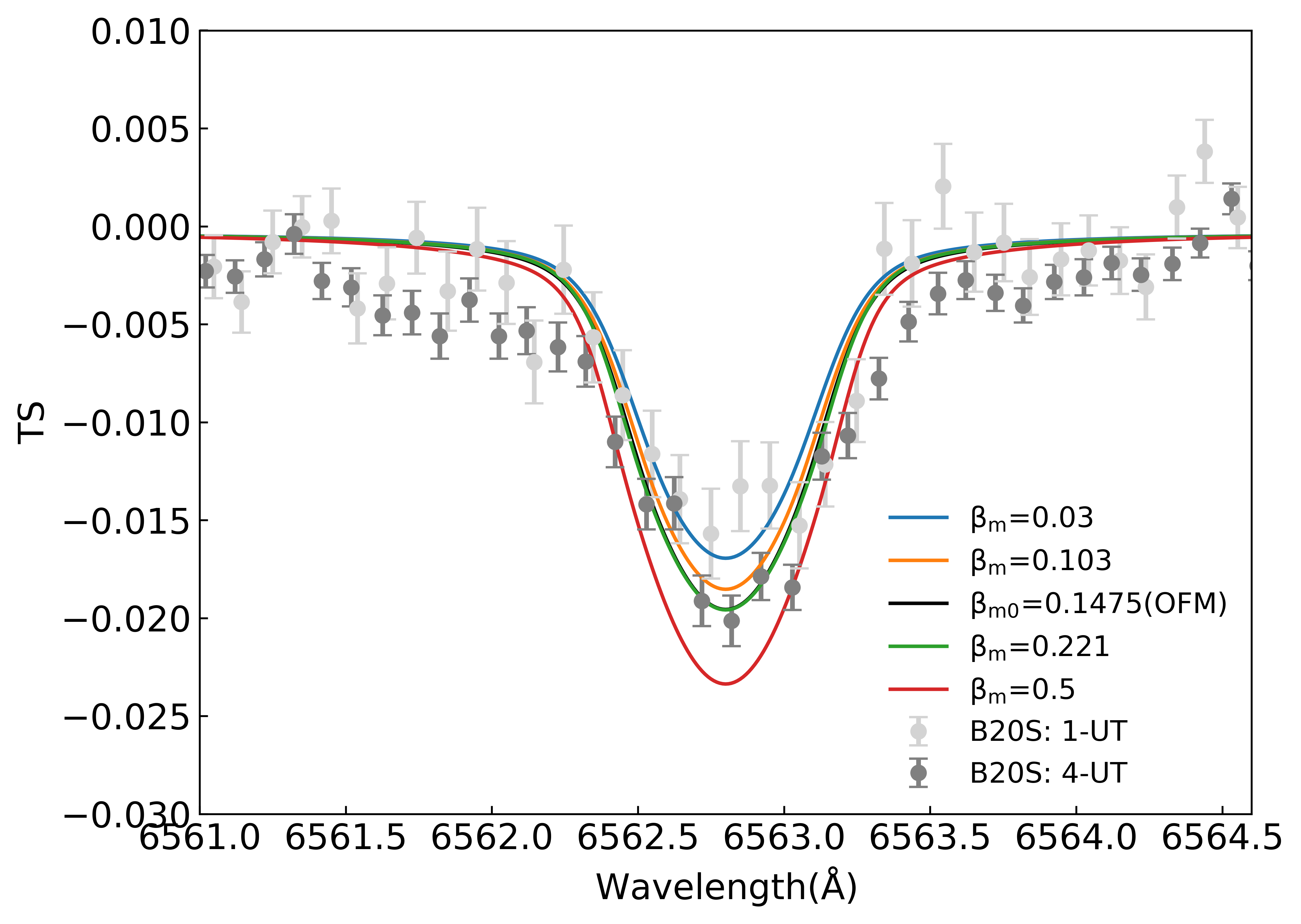}{0.5\textwidth}{(c)}
\fig{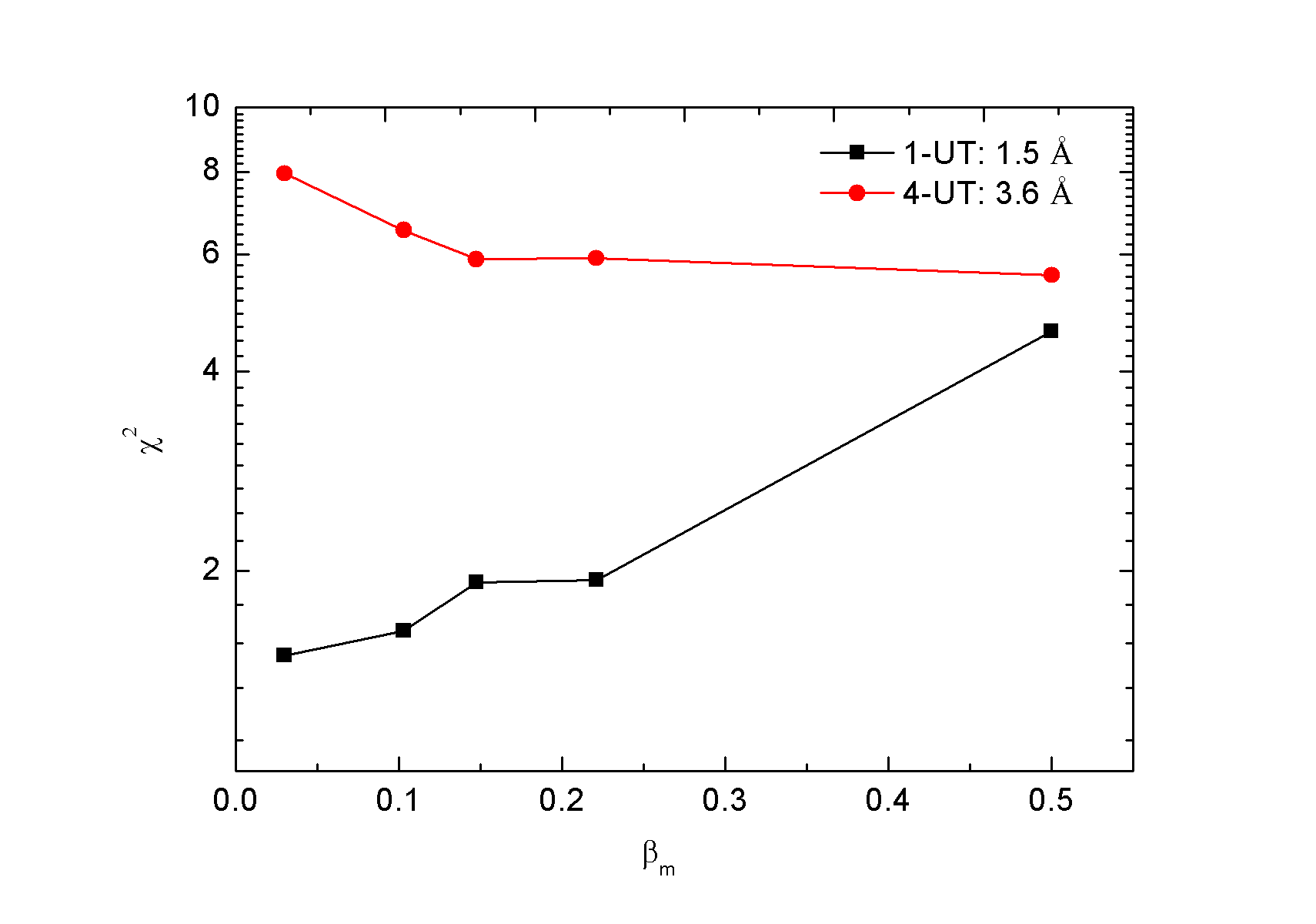}{0.6\textwidth}{(d)}
          }
\caption{\textbf{H$\alpha$ transmission spectrum and the comparison of models with observations.} (a) H$\alpha$ transmission spectrum. The gray dots with errors are the H$\alpha$ transmission spectrum of B20S, with the lightgray for 1-UT and darkgray for 4-UT.
Different lines represent models of different $F_{XUV}$, calculated within 7.6 R$_p$. (b) $\chi ^2$ as a function of different $F_{XUV}$. The black line represents the $\chi ^2$ for 1-UT, calculated in passband 1.5 $\rm\AA$.
The red line represents the $\chi ^2$ for 4-UT, calculated in passband 3.6 $\rm\AA$.
(c) The same as (a), but for the models of different XUV SEDs while $F_{XUV}$ = $F_0$, calculated within 7.6 R$_p$. Note that the black and green lines are almost overlapped. (d) The same as (b), but for the models of different XUV SEDs while $F_{XUV}$ = $F_0$.} \label{fidu_trans}
\end{figure*}

\begin{figure*}
\gridline{\fig{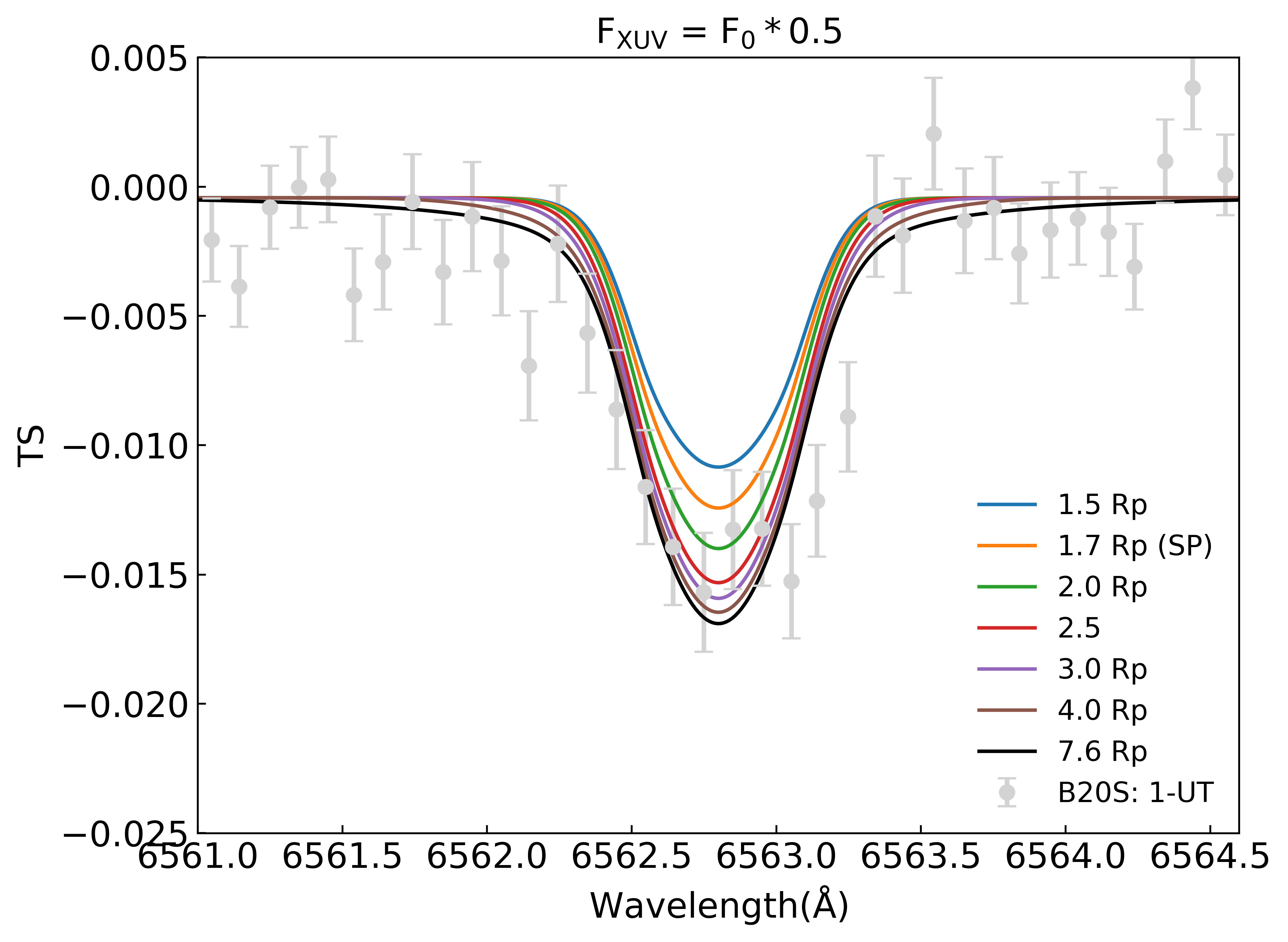}{0.5\textwidth}{(a)}
          \fig{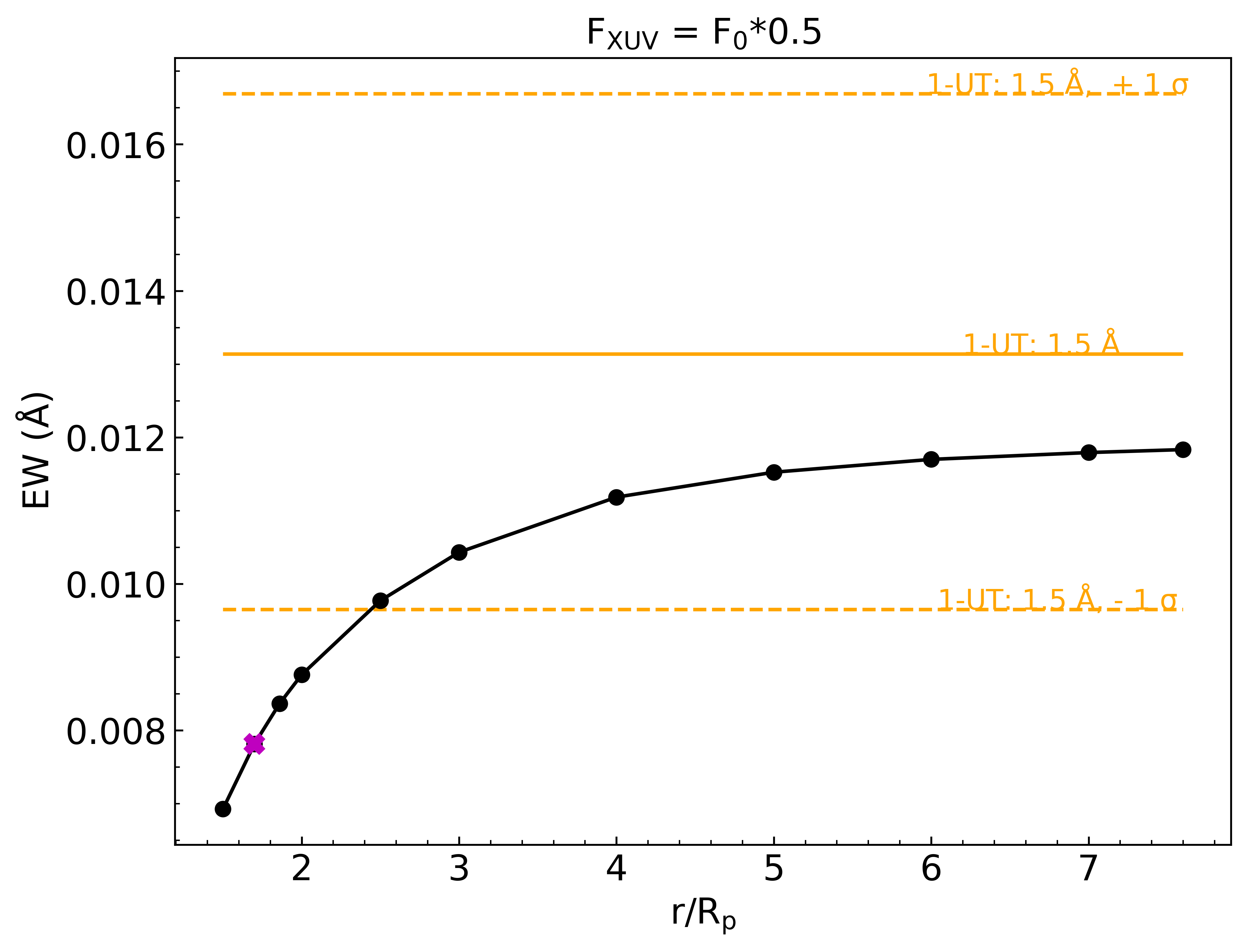}{0.5\textwidth}{(b)}
          }
\gridline{\fig{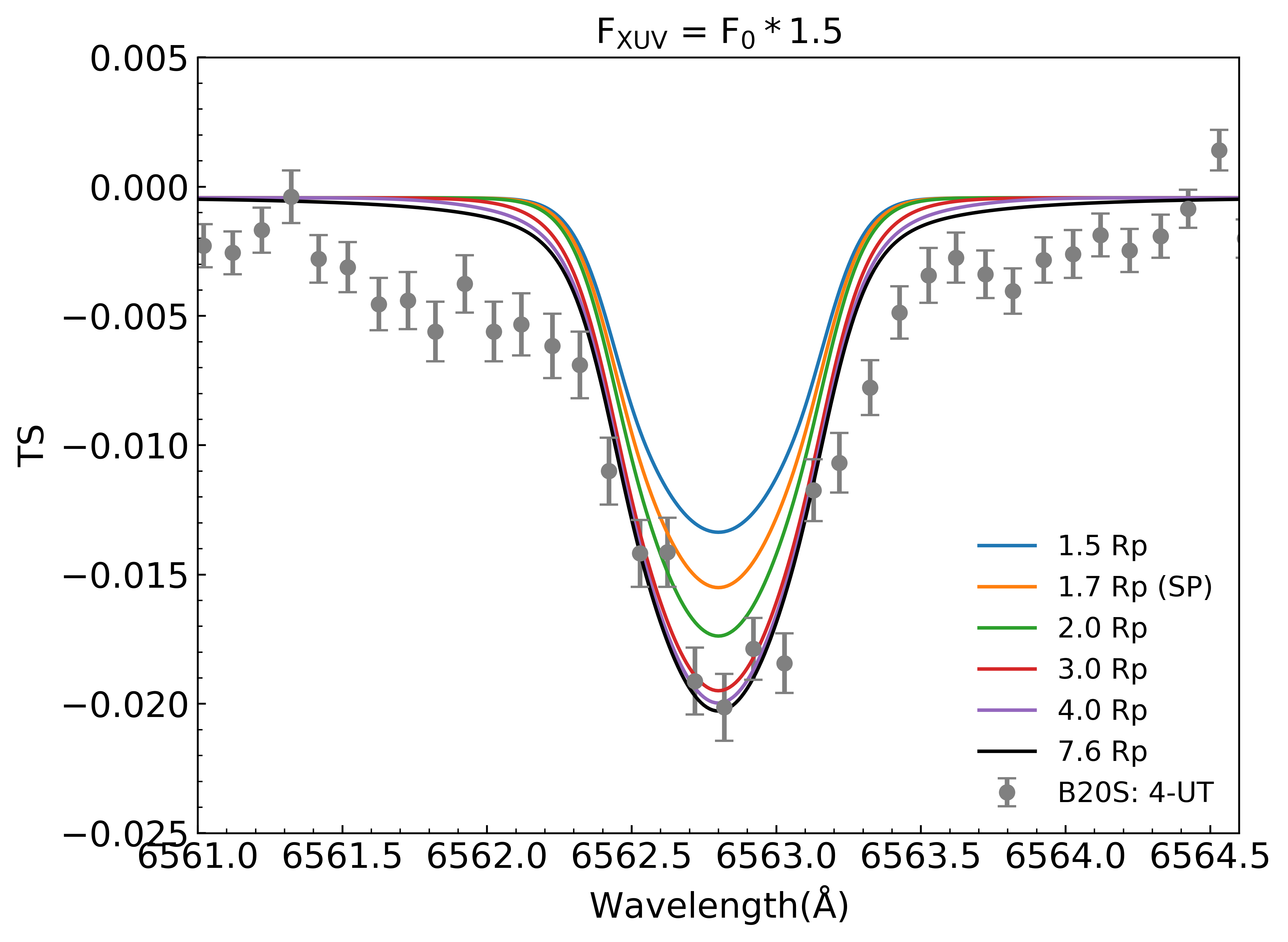}{0.5\textwidth}{(c)}
\fig{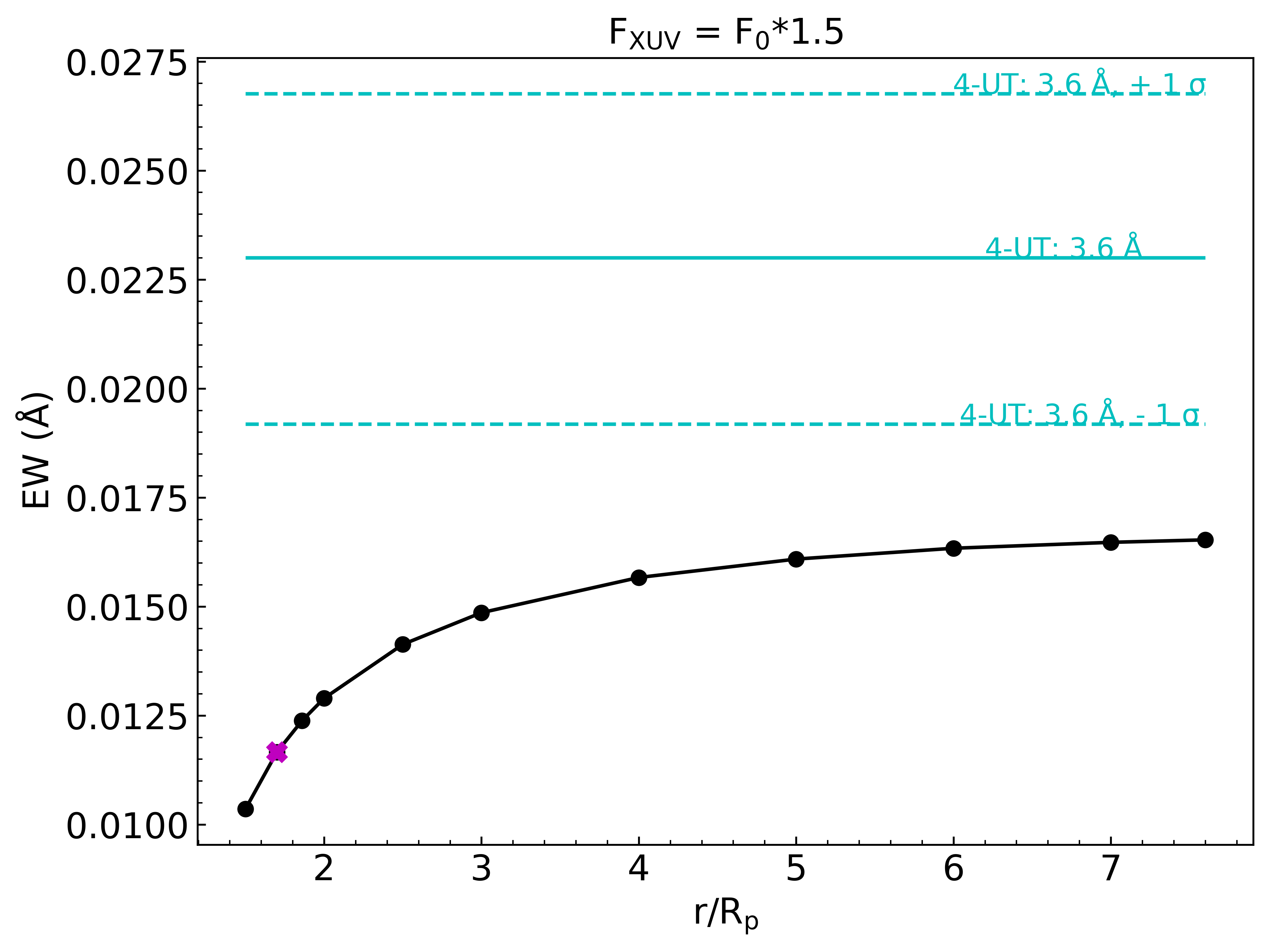}{0.5\textwidth}{(d)}
          }
\caption{\textbf{H$\alpha$ transmission spectrum and the comparison of models with observations for $F_{XUV}$=$F_0*0.5$ and $F_{XUV}$=$F_0*1.5$.} (a) H$\alpha$ transmission spectrum as a function of altitudes for $F_{XUV}$=$F_0*0.5$, in comparison with the observation of 1-UT. (b) The equivalent width calculated in passband 1.5 $\rm\AA$ as a function of atmospheric altitude, in comparison with the observation of 1-UT. The orange horizontal solid line represents the mean EW of 1-UT and the dashed lines are + 1 $\sigma$ and - 1 $\sigma$ of the mean EW. The sonic point is marked by the purple cross. (c) The same as (a), but for $F_{XUV}$=$F_0*1.5$ and 4-UT. (d) The same as (b), but for $F_{XUV}$=$F_0*1.5$ and 4-UT.}\label{sup_XUV}
\end{figure*}

\begin{figure*}
\gridline{\fig{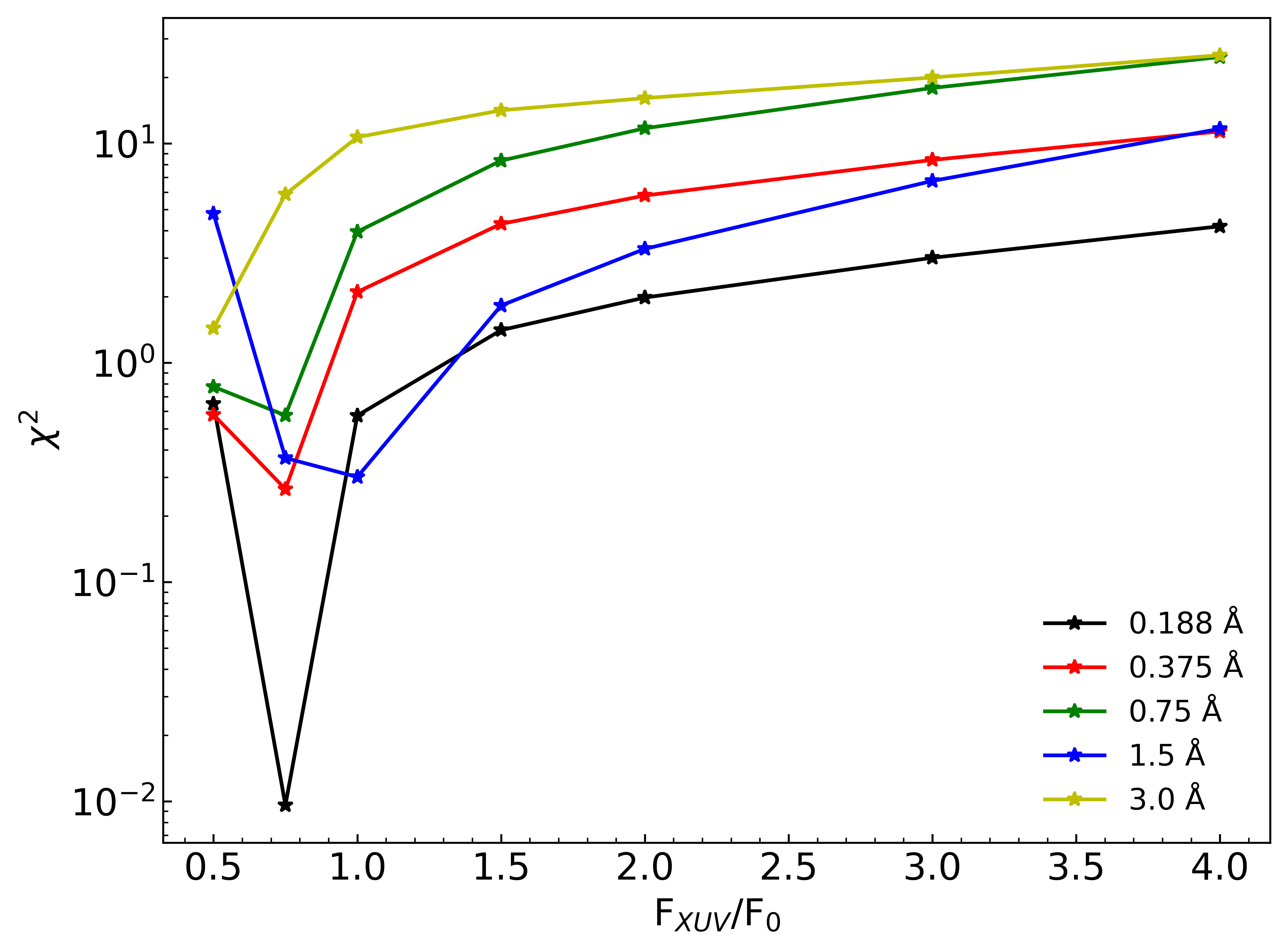}{0.5\textwidth}{(a)}
          }
\gridline{\fig{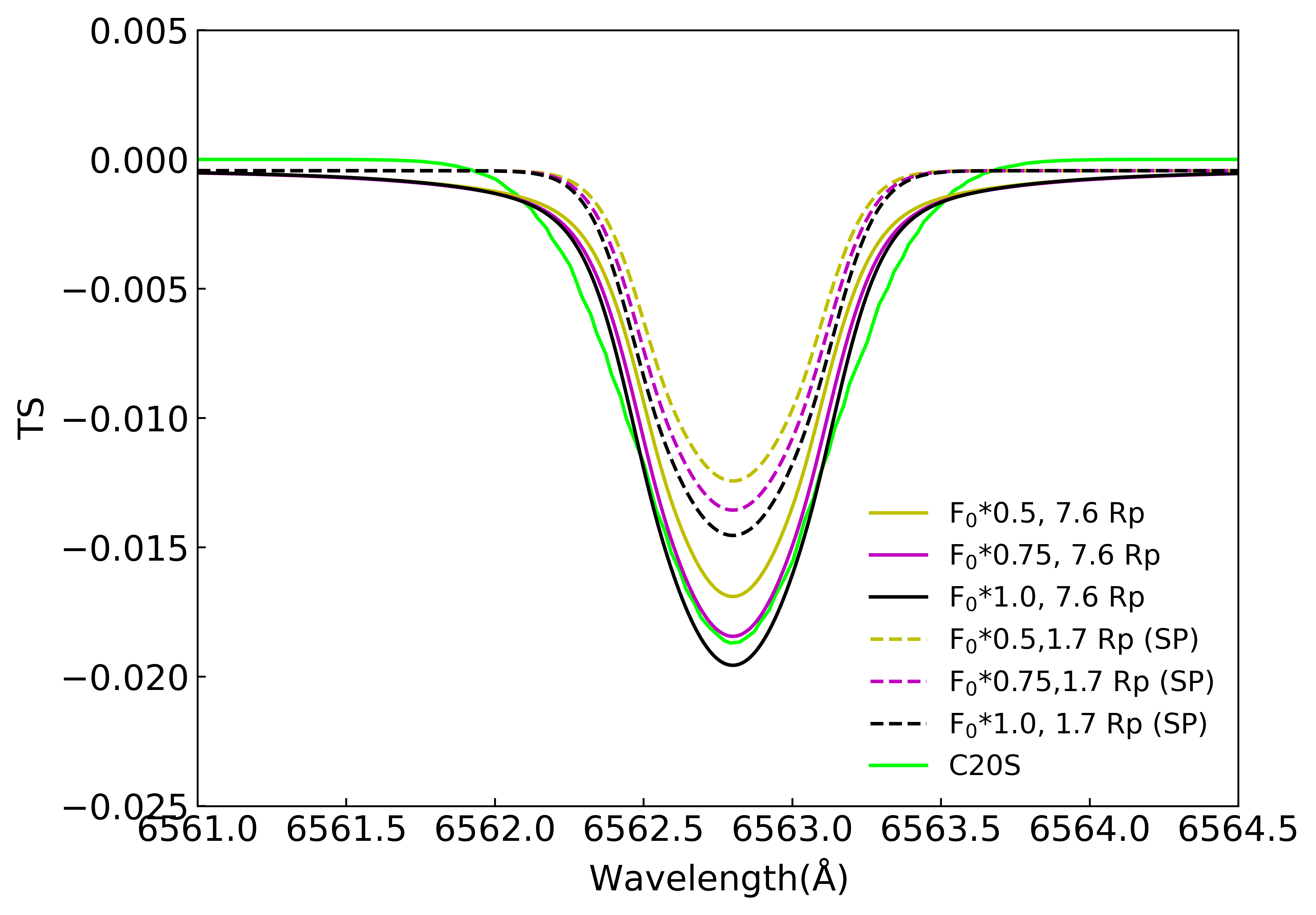}{0.55\textwidth}{(b)}
                    }
\gridline{\fig{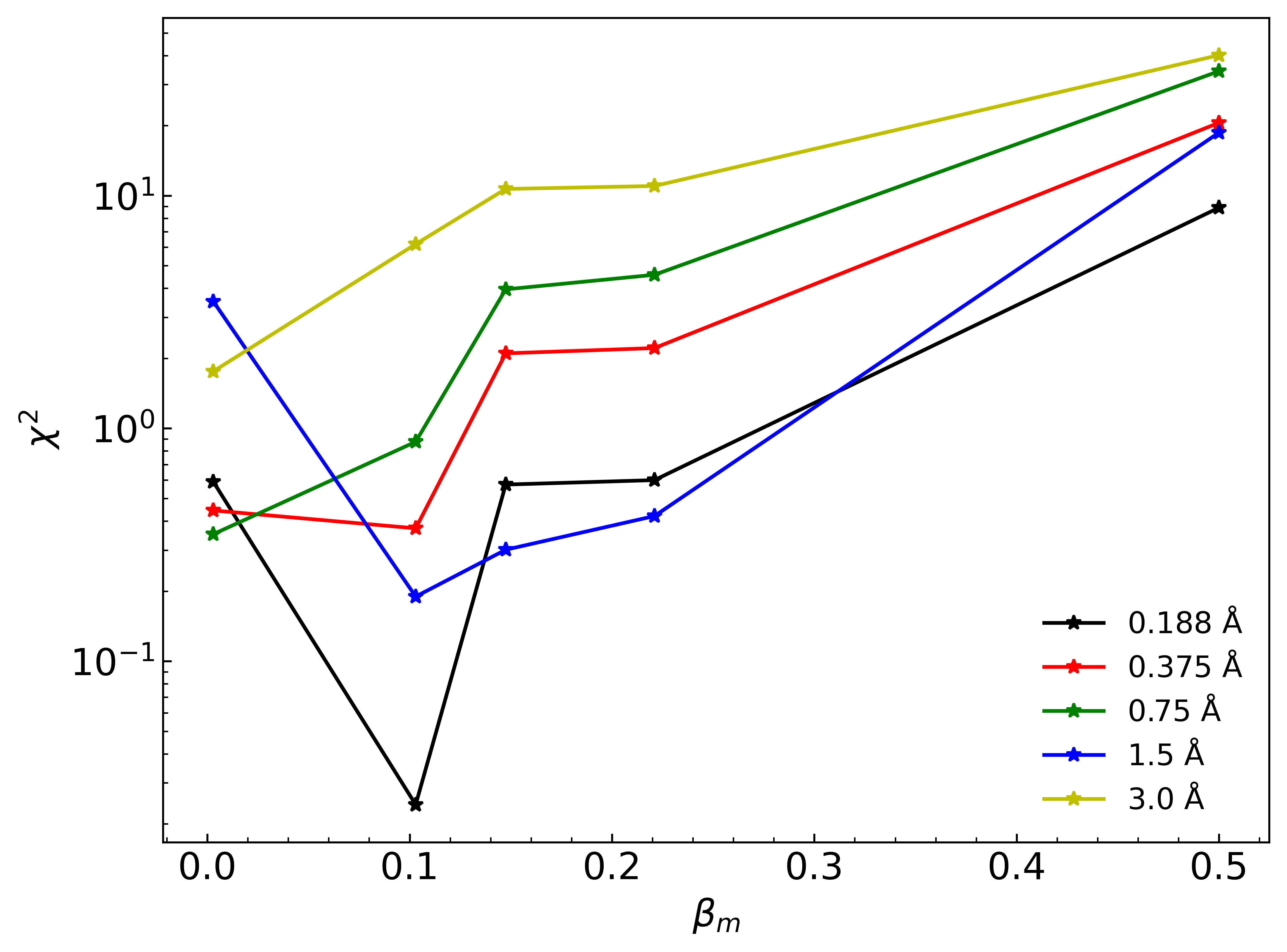}{0.5\textwidth}{(c)}
          }
\caption{\textbf{Comparison with C20.} (a) $\chi ^2$ as a function of different $F_{XUV}$. Different colors represent the $\chi ^2$ calculated by using different passbands. (b) Transmission spectrum for 0.5, 0.75 and 1.0 times $F_0$. The Green solid line represents C20S. Other solid lines are calculated within 7.6 R$_p$, and the dashed lines are calculated within the sonic points. (c) $\chi ^2$ as a function of different XUV SEDs, which are characterised by $\beta_m$.}\label{sup_XUV}
\end{figure*}

\end{document}